\documentclass[twoside,11pt]{article}

\usepackage[preprint]{jmlr2e}

\usepackage{amsmath}
\usepackage{bm}
\usepackage{booktabs}
\usepackage{multirow}
\usepackage{lastpage}

\usepackage{algorithm}
\usepackage{algorithmicx}
\usepackage{algpseudocode}

\newtheorem{assumption}{Assumption}

\jmlrheading{}{2026}{1-\pageref{LastPage}}{Submitted}{TBD}{TBD}{Lin, Zhou, Shi, and Sun}
\ShortHeadings{Projection-Based Partial Correlation Inference Beyond Sparsity}{Lin, Zhou, Shi, and Sun}
\firstpageno{1}

\begin{document}
\raggedbottom
\title{Projection-Based Partial Correlation Inference Beyond Sparsity}

\author{\name Youwu Lin
\email lyw015813@126.com \\
\addr Guanghua School of Management,\\ Peking University, Beijing 100871, P.R. China\\
and school of Mathematics and Computing Science,\\ Guilin University of Electronic Technology, \\Guilin, Guangxi 541002, P.R. China.
       \AND
       \name Zhaojun Zhou
       \email 13011760930@163.com\\
       \addr Linyi County Talent Development Service Center,\\ Linyi, Shandong 251518, P.R. China.
       \AND
       \name Shuqi Shi*
       \email shishuqi11@gmail.com\\
       \addr School of Finance, \\Zhongnan University of Economics and Law, \\Wuhan, Hubei 430073, P.R. China\\
       and school of Mathematics and Computing Science, \\Guilin University of Electronic Technology, Guilin,\\ Guangxi 541002, P.R. China
       \AND
       \name Xianming Sun
       \email xianming.sun@zuel.edu.cn\\
       \addr  School of Finance, \\
       and Innovation and Talent Base for Digital Technology and Finance,\\Zhongnan University of Economics and Law, \\ Wuhan, Hubei 430073, P.R. China\\
       $^*$ Corresponding author}

\maketitle

\begin{abstract}
Valid inference for partial correlation in high-dimensional Gaussian models is challenging when the conditioning variables exhibit dense dependence and conventional sparsity assumptions are inappropriate. We develop a target-specific projection framework for estimating and testing a population partial correlation without requiring coordinate sparsity. A population quadratic criterion characterizes the information retained by a linear reduction and is maximized by a one-dimensional projection. Under a rank-one alignment condition, this optimal projection preserves the original partial correlation. To obtain a feasible procedure, we estimate the optimal direction using diagonal ridge regularization on an independent training sample. Under conditions on regularized effective dimension and target-specific ridge bias, we establish a quantitative direction-estimation rate in the covariance-induced prediction norm. We further show that, under alignment, the induced partial-correlation error is second order in the direction error. These results yield oracle equivalence and asymptotic normality for the feasible split-sample estimator, as well as asymptotically valid testing of the original zero-partial-correlation null. Exact finite-sample \(F\)-calibration holds under the projected null. An explicit coordinate-dense Gaussian model verifies the required rates even when the dimension exceeds the sample size. Numerical experiments illustrate the finite-sample performance of the proposed procedures.
\end{abstract}


\section{Introduction}
\label{Sec:Introduction}

Partial correlation measures the linear association
between two variables after adjusting for the linear
effects of a set of covariates. Under joint Gaussianity,
zero partial correlation is equivalent to conditional
independence, making it a fundamental tool for
dependence analysis and statistical inference.
Applications include dependence network construction
in brain imaging \citep{luo2024functional} and
financial data analysis \citep{LW2017}.
Related developments have also considered more general
measures of conditional dependence
\citep{azadkia2021simple,huang2022kernel}.

In low-dimensional settings, partial correlation can
be estimated and tested using classical covariance-based
methods. However, with \(n\) observations and \(p\)
conditioning variables, the centered sample covariance
matrix is singular when \(p\ge n\), making the
conventional inverse-based estimator unavailable
\citep{liang2015equivalent}.
The underlying difficulty extends beyond matrix
singularity: although the partial correlation is a
scalar parameter, valid inference requires controlling
the effects of high-dimensional nuisance regression
components. Regularization can make the relevant
matrix inversions computationally feasible, but does
not by itself guarantee that these nuisance components
can be estimated accurately enough for inference.

A widely used approach addresses this difficulty by
imposing sparsity on regression coefficients or
precision-matrix components. Representative methods
include the graphical lasso
\citep{friedman_hastie_tibshirani_2008},
constrained \(\ell_1\)-minimization
\citep{cai2011constrained},
convex pseudolikelihood estimation
\citep{khare_oh_rajaratnam_2015},
and regression-based procedures
\citep{meinshausen_buhlmann_2006,peng2009,scalreg2013}.
Related applications of partial correlation to
variable selection have been studied under Gaussian
\citep{Buhlmann2010} and elliptical
\citep{Li2017} models.
For inference on precision-matrix entries,
\citet{ren2015asymptotic} established asymptotic
efficiency under suitable sparsity conditions.
These developments demonstrate the importance of
structural assumptions in high-dimensional inference.
Nevertheless, coordinate sparsity may be restrictive
when many correlated covariates jointly contribute
to the variables of interest. This raises the question
of whether alternative population structures can
support valid partial correlation inference without
requiring sparse regression or precision-matrix
representations.

We address this question through a target-specific
reduction framework for high-dimensional Gaussian
models. Rather than estimating the full covariance
or precision matrix, we seek a low-dimensional
linear projection of the conditioning variables
that preserves the original partial correlation.
Such a projection permits inference to be carried
out in a reduced model while retaining the
population parameter of interest. Related
projection-based approaches have been developed
for conditional dependence analysis, including
residual-based procedures
\citep{fan2020projection}.
Our framework focuses specifically on selecting
a projection that retains information relevant
to the original partial correlation.

To construct this reduction, we introduce a
population quadratic criterion measuring the
variation retained from a target-relevant
nuisance regression component. We characterize
its maximizers and show that the maximum is
attained by a one-dimensional direction.
Under a rank-one alignment condition, this
direction also preserves the original partial
correlation. We use the terms optimal projection
estimator (OPE) and optimal projection test (OPT)
for the resulting estimation and testing
procedures, where optimality refers to the
population retention criterion rather than
unrestricted estimation efficiency or testing
power. For a fixed target-preserving projection,
classical Gaussian theory yields asymptotically
normal estimation of the original partial
correlation and exact finite-sample central
\(F\) calibration under the zero-partial-correlation
null. These results provide an oracle benchmark,
but the population optimal direction is generally
unknown.

We therefore develop a feasible procedure based
on sample splitting and diagonal ridge
regularization. The projection direction is
estimated from a training sample of size \(n_1\),
while estimation and testing are conducted on
an independent evaluation sample of size \(n_2\).
To establish direction learnability, we derive
a quantitative error bound in the covariance-induced
prediction norm under conditions controlling
regularized effective dimension and target-specific
ridge bias. These conditions do not impose
coordinate sparsity, but require sufficient
population structure to estimate the relevant
directions. We verify the required rates in an
explicit coordinate-dense Gaussian model with
\(p\gg n\), demonstrating that the learning
conditions can hold even when the dimension
exceeds the sample size.

A key feature of the framework is the second-order
stability of the original target under rank-one
alignment. Specifically, the error in the
projected population partial correlation is
bounded by a quadratic function of the
prediction-norm direction error. Consequently,
a direction-estimation error of
\(o_p(n_2^{-1/4})\) makes the induced
population target error \(o_p(n_2^{-1/2})\).
Combined with independent evaluation and
control of the corresponding influence-function
difference, this yields asymptotic equivalence
between the feasible and oracle OPEs, together
with asymptotic normality for the original
partial correlation. For testing, the conditional
noncentral \(F\)-representation establishes
asymptotic validity under the original null.
Exact conditional central \(F\) calibration
holds under the projected null; for a learned
projection, validity under the original null
is generally asymptotic.

The resulting theory identifies a class of
structured dense high-dimensional models in
which valid partial correlation inference is
possible without coordinate sparsity.
Its guarantees rely on both target preservation
and sufficiently accurate direction learning,
and do not extend uniformly to unrestricted
dense covariance models. The framework thus
replaces coordinate sparsity with explicit
conditions on the population regression
structure and regularized learnability.

The remainder of the paper is organized as follows.
Section~\ref{s2} introduces the Gaussian model
and target-preserving projections.
Section~\ref{s3} develops the projected test,
characterizes the population optimal direction,
and establishes feasible inference.
Section~\ref{s4} presents numerical experiments.
Section~\ref{s5} discusses the scope and
limitations of the framework.
Proofs are provided in the appendices, with
additional numerical results in the
Supplementary Material.

\section{Model setup and target-preserving projections}
\label{s2}

\subsection{Model setup and the inferential target}

Let \(\bm W_1,\ldots,\bm W_n\) be independent
copies of
\[
\bm W=(Y,X,\bm Z^\top)^\top
\sim N_{p+2}(\bm 0,\bm\Sigma),
\qquad \bm\Sigma\succ0,
\]
where \(Y\) and \(X\) are scalar random
variables and \(\bm Z\in\mathbb R^p\)
is the conditioning vector.
In asymptotic analyses, both \(p=p_n\)
and \(\bm\Sigma=\bm\Sigma_n\) may vary
with \(n\).

Partition the covariance matrix according
to \((Y,X)^\top\) and \(\bm Z\):
\begin{equation}
\label{eq:Sigma_partition}
\bm\Sigma=
\begin{pmatrix}
\bm\Sigma_{11} & \bm\Sigma_{12}\\
\bm\Sigma_{21} & \bm\Sigma_{22}
\end{pmatrix},
\end{equation}
where
\(\bm\Sigma_{11}\in\mathbb R^{2\times2}\),
\(\bm\Sigma_{12}\in\mathbb R^{2\times p}\),
\(\bm\Sigma_{21}=\bm\Sigma_{12}^\top\),
and
\(\bm\Sigma_{22}
=\bm\Sigma_{\bm Z\bm Z}
=\operatorname{cov}(\bm Z)\).
For \(A\in\{Y,X\}\), define
\[
\bm\sigma_{\bm Z A}
=\operatorname{cov}(\bm Z,A),
\qquad
\bm\sigma_{A\bm Z}
=\bm\sigma_{\bm Z A}^\top.
\]

The population partial covariance matrix
of \((Y,X)^\top\) given \(\bm Z\) is
\[
\begin{aligned}
\bm\Sigma_{11\cdot2}
&=
\bm\Sigma_{11}
-
\bm\Sigma_{12}
\bm\Sigma_{22}^{-1}
\bm\Sigma_{21}\\
&=
\begin{pmatrix}
\sigma_{YY\cdot\bm Z}
&
\sigma_{YX\cdot\bm Z}\\
\sigma_{XY\cdot\bm Z}
&
\sigma_{XX\cdot\bm Z}
\end{pmatrix}.
\end{aligned}
\]
Under joint Gaussianity, this is also
the conditional covariance matrix of
\((Y,X)^\top\) given \(\bm Z\).
Our inferential target is the population
partial correlation
\begin{equation}
\label{eq:rho}
\rho=
\frac{\sigma_{YX\cdot\bm Z}}
{\sqrt{
\sigma_{YY\cdot\bm Z}
\sigma_{XX\cdot\bm Z}
}}.
\end{equation}
Since \(\bm\Sigma\succ0\), the residual
variances are positive and \(|\rho|<1\).

For an index set \(\mathcal I\) with
\(|\mathcal I|=m\ge2\), define the
sample mean and covariance matrix by
\[
\bar{\bm W}_{\mathcal I}
=
\frac1m\sum_{i\in\mathcal I}\bm W_i,
\qquad
\widehat{\bm\Sigma}_{\mathcal I}
=
\frac1{m-1}\sum_{i\in\mathcal I}
(\bm W_i-\bar{\bm W}_{\mathcal I})
(\bm W_i-\bar{\bm W}_{\mathcal I})^\top.
\]
All sample covariances are centered
and normalized by \(m-1\).
We partition
\(\widehat{\bm\Sigma}_{\mathcal I}\)
conformably with \(\bm\Sigma\) and omit
the subscript \(\mathcal I\) when no
ambiguity arises.
In particular,
\[
\widehat{\bm\sigma}_{\bm Z A}
=
\frac1{m-1}\sum_{i\in\mathcal I}
(\bm Z_i-\bar{\bm Z}_{\mathcal I})
(A_i-\bar A_{\mathcal I}),
\qquad
\widehat{\bm\sigma}_{A\bm Z}
=
\widehat{\bm\sigma}_{\bm Z A}^\top,
\]
for \(A\in\{Y,X\}\).
All sample partial correlations are
computed with an intercept.

When \(m>p+2\), the sample covariance
matrix is positive definite almost surely
under the Gaussian model. The sample
partial covariance matrix is therefore
\[
\widehat{\bm\Sigma}_{11\cdot2}
=
\widehat{\bm\Sigma}_{11}
-
\widehat{\bm\Sigma}_{12}
\widehat{\bm\Sigma}_{22}^{-1}
\widehat{\bm\Sigma}_{21},
\]
and the classical sample partial
correlation is
\begin{equation}
\label{eq:r}
r=
\frac{\widehat{\sigma}_{YX\cdot\bm Z}}
{\sqrt{
\widehat{\sigma}_{YY\cdot\bm Z}
\widehat{\sigma}_{XX\cdot\bm Z}
}}.
\end{equation}

In high dimensions, however,
\[
\operatorname{rank}
(\widehat{\bm\Sigma}_{22})
\le\min\{p,m-1\}.
\]
Thus \(\widehat{\bm\Sigma}_{22}\)
is singular whenever \(p\ge m\),
and the inverse-based estimator
in \eqref{eq:r} is undefined.
For the full sample, this occurs
when \(p\ge n\).
Although \(\rho\) is scalar,
its estimation requires adjustment
for high-dimensional covariate effects.
This motivates replacing \(\bm Z\)
with a low-dimensional projection
that preserves the original target
under explicit structural conditions.

\subsection{Target-preserving projections and oracle estimation}

We seek a low-dimensional linear projection of
\(\bm{Z}\) that preserves the original population
partial correlation. Such a reduction would allow
inference for \(\rho\) to be based on a
low-dimensional covariance matrix.

Let \(\bm{P}\in\mathbb{R}^{p\times t}\)
be a nonrandom full-column-rank matrix,
where \(1\leq t\leq p\), and define
\[
\bm{W}(\bm{P})
=
\bigl(Y,X,(\bm{P}^{\top}\bm{Z})^{\top}\bigr)^{\top}.
\]
Since \(\bm{\Sigma}\succ0\) and
\(\bm{P}\) has full column rank,
\(\bm{W}(\bm{P})\) follows a nondegenerate
Gaussian distribution.
Let \(\rho(\bm{P})\) denote the population
partial correlation between \(Y\) and \(X\)
given \(\bm{P}^{\top}\bm{Z}\).
Because dimension reduction generally changes
the conditioning information,
\(\rho(\bm{P})\) need not equal \(\rho\).
We therefore define target preservation
as follows.

\begin{definition}
\label{def:target-preserving}
A full-column-rank matrix
\(\bm{P}\in\mathbb{R}^{p\times t}\)
is called target-preserving if
\[
\rho(\bm{P})=\rho.
\]
\end{definition}

For an index set \(\mathcal{I}\) with
\(|\mathcal{I}|=m>t+2\), let
\(r_{\mathcal{I}}(\bm{P})\) denote the
sample partial correlation computed
from
\(\{\bm{W}_i(\bm{P}):i\in\mathcal{I}\}\),
including an intercept.
We write \(r(\bm{P})\) when the sample
index set is clear from the context.
Under the Gaussian model, the projected
sample covariance matrix is positive
definite almost surely.
Thus \(r_{\mathcal{I}}(\bm{P})\)
is well defined and estimates
\(\rho(\bm{P})\).
When \(\bm{P}\) is target-preserving,
inference for \(\rho(\bm{P})\) also
applies to the original target \(\rho\).

\begin{theorem}
\label{thm:ope_asymp}
Suppose that the Gaussian model holds
and that
\(\bm{P}\in\mathbb{R}^{p\times t}\)
is a nonrandom target-preserving
projection, where \(t\) is fixed.
If \(|\rho|\leq1-\delta\) for some
constant \(\delta>0\), then, as
\(n\to\infty\),
\[
\frac{
\sqrt{n}\{r(\bm{P})-\rho\}
}{
1-\rho^2
}
\xrightarrow{d}N(0,1).
\]
\end{theorem}

Theorem~\ref{thm:ope_asymp}
provides an oracle benchmark for
inference based on a nonrandom
target-preserving projection.
The remaining challenge is to identify
such a projection and estimate it
without introducing an error that
affects inference for the original target.

In Section~\ref{s3}, we introduce a
population retention criterion and
characterize its optimal one-dimensional
direction \(\bm{P}^{\ast}\).
Proposition~\ref{prop:target_preserving_pstar}
establishes target preservation under
rank-one alignment.
The resulting oracle optimal projection
estimator (OPE) provides a benchmark for
the feasible procedure, which learns the
direction from a training sample and
performs inference on an independent
evaluation sample. Validity of the
feasible estimator requires controlling
the error induced by replacing
\(\bm{P}^{\ast}\) with its estimate.

\section{Hypothesis testing under the reduction framework}
\label{s3}

Section~\ref{s2} establishes that a suitable
linear projection can preserve the original
partial correlation while reducing the
conditioning dimension. We now develop
the corresponding testing framework for
\[
H_0:\rho=0
\qquad\text{versus}\qquad
H_1:\rho\ne0.
\]

For a fixed projection \(\bm P\), the
projected variables remain jointly Gaussian,
so classical partial-correlation testing
theory applies. The resulting \(F\)-statistic
has an exact central \(F\)-distribution
under the projected null
\(\rho(\bm P)=0\) and a conditional
noncentral \(F\)-representation under
the projected alternative.
When \(\bm P\) is target-preserving,
the projected and original nulls coincide.
For a direction estimated from independent
training data, exact conditional calibration
still concerns the projected null, whereas
validity for the original null requires
control of the projection-induced target error.
We establish these distributional results
before developing population projection
selection and feasible inference.

\subsection{The \(F\)-test in low-dimensional settings}

We first consider the classical Gaussian
setting with \(n>p+2\), where the sample
partial correlation \(r\) in
\eqref{eq:r} is well defined.
Testing \(H_0:\rho=0\) is equivalent
to testing whether the coefficient
of \(X\) vanishes in the linear
regression of \(Y\) on \(X\) and
\(\bm Z\), including an intercept.
The corresponding \(F\)-statistic is
\begin{equation}
\label{eq:F}
F
=
(n-p-2)\frac{r^2}{1-r^2}.
\end{equation}

To characterize its distribution under
the alternative, let
\[
a_{XX\cdot\bm Z}
=
\sum_{i=1}^n
(X_i-\widehat X_i)^2
=
(n-1)\widehat{\sigma}_{XX\cdot\bm Z},
\]
where \(\widehat X_i\) is the fitted
value from regressing \(X\) on
an intercept and \(\bm Z\).
The second equality follows from
the covariance normalization in
Section~\ref{s2}.

The following lemma summarizes the
distributional properties needed for
the projected test. We refer to
\citet[Chapter~5]{muirhead1982aspects}
for the classical Gaussian theory
and provide a proof in the appendix.
Let \(F_{d_1,d_2}(\delta)\) denote the
noncentral \(F\)-distribution with
degrees of freedom \(d_1,d_2\)
and noncentrality parameter \(\delta\).

\begin{lemma}
\label{lem:low_dim_F}
Suppose that \(n\) independent
observations follow the Gaussian
model in Section~\ref{s2}, with
\(n>p+2\).
Let \(r\) be the sample partial
correlation between \(Y\) and \(X\)
given \(\bm Z\), and define \(F\)
by \eqref{eq:F}.
Then
\[
\frac{a_{XX\cdot\bm Z}}
{\sigma_{XX\cdot\bm Z}}
\sim\chi^2_{n-p-1}.
\]
Moreover, the following statements hold:
\begin{enumerate}
\item Under \(H_0:\rho=0\),
\[
F\sim F_{1,n-p-2}.
\]

\item Under \(H_1:\rho\ne0\),
\[
F\mid a_{XX\cdot\bm Z}
\sim
F_{1,n-p-2}
\left\{
\delta_c(a_{XX\cdot\bm Z})
\right\},
\]
where
\[
\delta_c(a)
=
\frac{a}{\sigma_{XX\cdot\bm Z}}
\frac{\rho^2}{1-\rho^2}.
\]
The unconditional alternative
distribution is obtained by mixing
these conditional noncentral
\(F\)-distributions over the
distribution of \(a_{XX\cdot\bm Z}\).
\end{enumerate}
\end{lemma}

For any fixed full-column-rank
matrix
\(\bm P\in\mathbb R^{p\times t}\),
Lemma~\ref{lem:low_dim_F}
applies with \(\bm Z\) replaced by
\(\bm P^\top\bm Z\) and \(p\)
replaced by \(t\), provided that
\(n>t+2\). This establishes the
distributional basis for the
projected \(F\)-test in the next
subsection, whose exact null
calibration concerns
\(\rho(\bm P)=0\).

\subsection{Reduction-based testing in high dimensions}

We extend the classical \(F\)-test to
projected observations, allowing the
ambient dimension \(p\) to exceed the
sample size. Let
\(\bm P\in\mathbb R^{p\times t}\)
be a nonrandom full-column-rank matrix,
where \(1\le t\le p\) and \(n>t+2\),
and define
\[
\bm U_{\bm P}=\bm P^\top\bm Z.
\]
For \(A,B\in\{Y,X\}\), let
\(\sigma_{AB\cdot\bm P}\) denote the
population residual covariance after
adjusting for \(\bm U_{\bm P}\).
The projected population partial
correlation is
\[
\rho(\bm P)
=
\frac{\sigma_{YX\cdot\bm P}}
{\sqrt{
\sigma_{YY\cdot\bm P}
\sigma_{XX\cdot\bm P}
}}.
\]

Let \(r(\bm P)\) denote the sample
partial correlation between \(Y\)
and \(X\) given \(\bm U_{\bm P}\),
computed with an intercept.
Define the projected residual sum
of squares by
\[
a_{XX\cdot\bm P}
=
\sum_{i=1}^n
\{X_i-\widehat X_i(\bm P)\}^2
=
(n-1)\widehat{\sigma}_{XX\cdot\bm P},
\]
where \(\widehat X_i(\bm P)\)
is the fitted value from regressing
\(X\) on an intercept and
\(\bm P^\top\bm Z\).

Since \(\bm P\) is nonrandom and
has full column rank, the projected
variables remain jointly Gaussian.
Lemma~\ref{lem:low_dim_F} therefore
yields the following result.

\begin{theorem}
\label{thm:splitvalid}
Suppose that \(n\) independent
observations follow the Gaussian
model in Section~\ref{s2}.
Let
\(\bm P\in\mathbb R^{p\times t}\)
be a nonrandom full-column-rank
matrix, where \(1\le t\le p\)
and \(n>t+2\).
Define
\[
F(\bm P)
=
(n-t-2)
\frac{r^2(\bm P)}
{1-r^2(\bm P)}.
\]
Then
\[
\frac{a_{XX\cdot\bm P}}
{\sigma_{XX\cdot\bm P}}
\sim\chi^2_{n-t-1}.
\]
Moreover, the following statements
hold:
\begin{enumerate}

\item Under the projected null
\(H_{0,\bm P}:\rho(\bm P)=0\),
\[
F(\bm P)\sim F_{1,n-t-2}.
\]

\item Under the projected alternative
\(H_{1,\bm P}:\rho(\bm P)\ne0\),
\[
F(\bm P)\mid a_{XX\cdot\bm P}
\sim
F_{1,n-t-2}
\left\{
\delta_c(\bm P,a_{XX\cdot\bm P})
\right\},
\]
where
\[
\begin{aligned}
\delta_c(\bm P,a)
&=
\frac{a}{\sigma_{XX\cdot\bm P}}
\frac{\rho^2(\bm P)}
{1-\rho^2(\bm P)}\\
&=
\frac{
a\,\sigma_{YX\cdot\bm P}^2
}{
\sigma_{XX\cdot\bm P}^2
\left(
\sigma_{YY\cdot\bm P}
-
\frac{\sigma_{YX\cdot\bm P}^2}
{\sigma_{XX\cdot\bm P}}
\right)
}.
\end{aligned}
\]
The unconditional alternative
distribution is obtained by
mixing these conditional noncentral
\(F\)-distributions over the
distribution of \(a_{XX\cdot\bm P}\).

\end{enumerate}
\end{theorem}

Theorem~\ref{thm:splitvalid}
establishes exact finite-sample
calibration under the projected
null \(H_{0,\bm P}\).
If \(\bm P\) is target-preserving,
then \(\rho(\bm P)=\rho\),
and the same calibration applies
to the original null \(H_0:\rho=0\).
Moreover, since
\[
\frac{a_{XX\cdot\bm P}}
{\sigma_{XX\cdot\bm P}}
\sim\chi^2_{n-t-1},
\]
the unconditional distribution
of \(F(\bm P)\), for fixed \(t\),
depends on \(\bm P\) only through
\(\rho^2(\bm P)\).

The remaining challenge is to
select a low-dimensional projection
that preserves the original target.
In the next subsection, we introduce
a population retention criterion,
characterize its maximizers, and
establish a rank-one alignment
condition under which an optimal
one-dimensional direction is
target-preserving. Optimality refers
to this retention criterion.

\subsection{Population projection selection and target preservation}
\label{sec_optimal_projection}

Theorem~\ref{thm:splitvalid} establishes the
distributional properties of the projected
\(F\)-test for a fixed projection.
To construct a reduction suitable for
inference on the original partial correlation,
we seek a low-dimensional projection that
retains the relevant covariate information
and preserves the population target.
We first introduce a population retention
criterion and characterize its maximizers,
then establish a condition under which
an optimal one-dimensional direction
is target-preserving.

Refine the covariance partition in
\eqref{eq:Sigma_partition} as
\begin{equation}
\label{eq:fine_partition}
\bm{\Sigma}
=
\begin{pmatrix}
\sigma_{YY}
&
\sigma_{YX}
&
\bm{\sigma}_{Y\bm{Z}}
\\
\sigma_{XY}
&
\sigma_{XX}
&
\bm{\sigma}_{X\bm{Z}}
\\
\bm{\sigma}_{\bm{Z}Y}
&
\bm{\sigma}_{\bm{Z}X}
&
\bm{\Sigma}_{\bm{Z}\bm{Z}}
\end{pmatrix},
\end{equation}
where
\(\bm{\sigma}_{\bm{Z}A}
=\operatorname{cov}(\bm{Z},A)\)
and
\(\bm{\sigma}_{A\bm{Z}}
=\bm{\sigma}_{\bm{Z}A}^{\top}\)
for \(A\in\{Y,X\}\).

Define
\[
\gamma
=
\frac{\sigma_{YX\cdot\bm{Z}}}
{\sigma_{XX\cdot\bm{Z}}},
\qquad
Y_{\gamma}=Y-\gamma X,
\]
and
\[
\bm{\xi}
=
\operatorname{cov}(\bm{Z},Y_{\gamma})
=
\bm{\sigma}_{\bm{Z}Y}
-
\gamma\bm{\sigma}_{\bm{Z}X}.
\]
Here \(\gamma\) is the coefficient of
\(X\) in the population regression of
\(Y\) on \(X\) and \(\bm{Z}\).
Under the Gaussian model,
\[
\mathbb{E}(Y_{\gamma}\mid\bm{Z})
=
\bm{v}^{\top}\bm{Z},
\qquad
\bm{v}
=
\bm{\Sigma}_{\bm{Z}\bm{Z}}^{-1}
\bm{\xi}.
\]

For a full-column-rank matrix
\(\bm{P}\in\mathbb{R}^{p\times t}\),
define the population retention criterion
\[
\begin{aligned}
Q(\bm{P})
&:=
\operatorname{var}
\left\{
\mathbb{E}
(Y_{\gamma}\mid\bm{P}^{\top}\bm{Z})
\right\}\\
&=
\bm{\xi}^{\top}\bm{P}
\left(
\bm{P}^{\top}
\bm{\Sigma}_{\bm{Z}\bm{Z}}
\bm{P}
\right)^{-1}
\bm{P}^{\top}\bm{\xi}.
\end{aligned}
\]
This criterion measures the variation
in the target-relevant regression
component retained by the projection.
The corresponding variation under
the full conditioning vector is
\[
q
:=
\operatorname{var}
\left\{
\mathbb{E}(Y_{\gamma}\mid\bm{Z})
\right\}
=
\bm{\xi}^{\top}
\bm{\Sigma}_{\bm{Z}\bm{Z}}^{-1}
\bm{\xi}.
\]

\begin{theorem}
\label{thm:optproj}
Suppose that the Gaussian model in
Section~\ref{s2} holds and \(q>0\).
For every full-column-rank matrix
\(\bm{P}\in\mathbb{R}^{p\times t}\),
where \(1\leq t\leq p\),
\[
0\leq Q(\bm{P})\leq q,
\]
with
\[
Q(\bm{P})=q
\quad\Longleftrightarrow\quad
\bm{v}\in\operatorname{col}(\bm{P}).
\]
In particular, the one-dimensional
direction
\[
\bm{P}^{\ast}
=
\frac{\bm{v}}{\sqrt q}
=
\frac{
\bm{\Sigma}_{\bm{Z}\bm{Z}}^{-1}
\bm{\xi}
}{\sqrt q}
\]
satisfies
\[
(\bm{P}^{\ast})^{\top}
\bm{\Sigma}_{\bm{Z}\bm{Z}}
\bm{P}^{\ast}
=1,
\qquad
Q(\bm{P}^{\ast})=q.
\]
Thus the maximum retention is
attained with projected dimension one.
\end{theorem}

Theorem~\ref{thm:optproj} shows that
the entire regression component
can be retained by a single projected
variable. Indeed, letting
\[
T=(\bm{P}^{\ast})^{\top}\bm{Z},
\]
we have
\[
\mathbb{E}(Y_{\gamma}\mid T)
=
\mathbb{E}(Y_{\gamma}\mid\bm{Z})
=
\sqrt q\,T
\qquad\text{almost surely}.
\]
We therefore select the
lowest-dimensional maximizer of
\(Q\). For \(n>3\), this choice also
provides the largest denominator
degrees of freedom, \(n-3\), among
the projected \(F\)-tests with
dimensions \(t\geq1\).

Maximizing \(Q\), however, does not
by itself guarantee preservation of
the original partial correlation.
The following proposition provides
a sufficient condition.

\begin{proposition}
\label{prop:target_preserving_pstar}
Suppose that the conditions of
Theorem~\ref{thm:optproj} hold and
\[
\bm{\Sigma}_{\bm{Z}\bm{Z}}^{-1}
\bm{\sigma}_{\bm{Z}X}
\in
\operatorname{col}(\bm{P}^{\ast}).
\]
Then
\[
\begin{aligned}
\sigma_{YY\cdot\bm{P}^{\ast}}
&=
\sigma_{YY\cdot\bm{Z}},\\
\sigma_{XX\cdot\bm{P}^{\ast}}
&=
\sigma_{XX\cdot\bm{Z}},\\
\sigma_{YX\cdot\bm{P}^{\ast}}
&=
\sigma_{YX\cdot\bm{Z}}.
\end{aligned}
\]
Consequently,
\[
\rho(\bm{P}^{\ast})=\rho.
\]
\end{proposition}

\begin{remark}
The alignment condition admits
a regression interpretation.
Write
\[
X=\bm{\beta}^{\top}\bm{Z}
+\varepsilon_2,
\qquad
Y=\bm{\alpha}^{\top}\bm{Z}
+\varepsilon_1,
\]
where
\[
\bm{\beta}
=
\bm{\Sigma}_{\bm{Z}\bm{Z}}^{-1}
\bm{\sigma}_{\bm{Z}X},
\qquad
\bm{\alpha}
=
\bm{\Sigma}_{\bm{Z}\bm{Z}}^{-1}
\bm{\sigma}_{\bm{Z}Y}.
\]
Under joint Gaussianity,
\(\bm{\varepsilon}
=(\varepsilon_1,\varepsilon_2)^\top\)
is independent of \(\bm{Z}\), and
\[
\bm{v}
=
\bm{\alpha}-\gamma\bm{\beta}.
\]
Since \(q>0\), the alignment
condition is equivalent to
\(\bm{\alpha}\) and \(\bm{\beta}\)
lying in a common one-dimensional
subspace. In particular, if
\(\bm{\beta}=c\bm{P}^{\ast}\)
for some \(c\in\mathbb{R}\), then
\[
\bm{\alpha}
=
(\sqrt q+\gamma c)\bm{P}^{\ast}.
\]
Thus both population regression
components depend on \(\bm{Z}\)
only through \(T\), yielding
the residual covariance preservation
in Proposition~%
\ref{prop:target_preserving_pstar}.

Alignment permits coordinate-dense
regression coefficients but restricts
their joint regression space to
dimension one. It is sufficient for
target preservation, whereas maximizing
\(Q\) does not require alignment.
More generally, the span of
\(\bm{\alpha}\) and \(\bm{\beta}\),
of dimension at most two, preserves
the full residual covariance matrix.

For the designs satisfying
\(\bm{\beta}=-\bm{\alpha}\)
in Section~\ref{s4},
\[
\bm{v}=(1+\gamma)\bm{\alpha}.
\]
Hence alignment holds whenever
\((1+\gamma)\bm{\alpha}\ne\bm{0}\).
This verifies population target
preservation; learnability of the
direction requires separate conditions.
\end{remark}

Under the conditions of
Proposition~\ref{prop:target_preserving_pstar},
the projected and original nulls
coincide. Thus, for \(n>3\),
Theorem~\ref{thm:splitvalid}
gives
\[
F(\bm{P}^{\ast})
\sim F_{1,n-3}
\qquad\text{under }H_0:\rho=0.
\]
This is an oracle result because
\(\bm{P}^{\ast}\) depends on unknown
population quantities.
The next subsection develops a
feasible ridge-based estimator
of this direction and quantifies
the resulting direction and
target errors.

\subsection{Feasible reduction and inferential implementation}

The population optimal direction in
Theorem~\ref{thm:optproj} depends on unknown
covariance quantities. To obtain a feasible
procedure, we estimate this direction from
a training sample and conduct inference on
an independent evaluation sample.
Partition the observation indices,
independently of the observed values,
into \(\mathcal I_1\) and \(\mathcal I_2\), with
\[
|\mathcal I_1|=n_1\ge3,
\qquad
|\mathcal I_2|=n_2>3,
\qquad
n_1+n_2=n.
\]
All training-sample covariances are centered
and normalized by \(1/(n_1-1)\), as specified
in Section~\ref{s2}.

Since the training covariance matrix of
\(\bm Z\) is singular when \(p\ge n_1\),
we use diagonal ridge regularization.
Related ridge-based projection estimators
have been considered by
\citet{li2022linear}.
Define
\[
\widehat{\bm D}_{\bm Z}
=
\operatorname{diag}
(\widehat{\bm\Sigma}_{\bm Z\bm Z}),
\qquad
\widehat{\bm A}
=
\widehat{\bm\Sigma}_{\bm Z\bm Z}
+
\lambda_n\widehat{\bm D}_{\bm Z},
\qquad
\widehat{\bm M}
=
\widehat{\bm A}^{-1},
\]
where \(\lambda_n>0\) is deterministic.
Under the Gaussian model,
\(\widehat{\bm D}_{\bm Z}\succ0\)
almost surely, so
\(\widehat{\bm A}\succ0\)
almost surely.
This regularization corresponds to
an identity ridge penalty after
standardizing the covariates by their
training-sample standard deviations.

Using \(\widehat{\bm M}\), define
the ridge-adjusted covariance quantities
\[
\begin{aligned}
\widehat{\sigma}_{YX\cdot\bm Z}^{\,\mathrm{ridge}}
&=
\widehat{\sigma}_{YX}
-
\widehat{\bm\sigma}_{Y\bm Z}
\widehat{\bm M}
\widehat{\bm\sigma}_{\bm Z X},\\
\widehat{\sigma}_{XX\cdot\bm Z}^{\,\mathrm{ridge}}
&=
\widehat{\sigma}_{XX}
-
\widehat{\bm\sigma}_{X\bm Z}
\widehat{\bm M}
\widehat{\bm\sigma}_{\bm Z X}.
\end{aligned}
\]
The second quantity is positive almost
surely. Indeed, the empirical covariance
matrix of \((X,\bm Z^\top)^\top\) is
positive semidefinite, with
\(\widehat{\sigma}_{XX}>0\) and
\(\widehat{\bm D}_{\bm Z}\succ0\)
almost surely. Adding a positive
ridge penalty to its covariate block
gives a positive-definite matrix,
whose Schur complement with respect
to \(\widehat{\bm A}\) is positive.

Consequently, define
\[
\widehat{\gamma}
=
\frac{
\widehat{\sigma}_{YX\cdot\bm Z}^{\,\mathrm{ridge}}
}{
\widehat{\sigma}_{XX\cdot\bm Z}^{\,\mathrm{ridge}}
},
\qquad
\widehat{\bm\xi}
=
\widehat{\bm\sigma}_{\bm Z Y}
-
\widehat{\gamma}
\widehat{\bm\sigma}_{\bm Z X},
\]
and
\[
\widehat q
=
\widehat{\bm\xi}^{\top}
\widehat{\bm M}
\widehat{\bm\xi}
\ge0.
\]
The feasible ridge direction is
\begin{equation}
\label{eq:P1_opt_Ri}
\widehat{\bm P}_{\mathrm{ridge}}^\ast
=
\begin{cases}
\displaystyle
\frac{
\widehat{\bm M}\widehat{\bm\xi}
}{
\sqrt{\widehat q}
},
& \widehat q>0,\\[1.2ex]
\bm e_1,
& \widehat q=0,
\end{cases}
\end{equation}
where \(\bm e_1\) is the first
coordinate vector in \(\mathbb R^p\).
The convention for \(\widehat q=0\)
ensures that the direction is nonzero.

Inference is subsequently performed
on the evaluation sample using the
ordinary partial correlation between
\(Y\) and \(X\) after adjusting for
\((\widehat{\bm P}_{\mathrm{ridge}}^\ast)^\top
\bm Z\), including an intercept.
Validity for the original parameter
requires control of both the direction
estimation error and its effect on
the population partial correlation.

To quantify these errors, define the
population marginal-variance matrix
and correlation matrix by
\[
\bm D_{\bm Z}
=
\operatorname{diag}
(\bm\Sigma_{\bm Z\bm Z}),
\qquad
\bm R
=
\bm D_{\bm Z}^{-1/2}
\bm\Sigma_{\bm Z\bm Z}
\bm D_{\bm Z}^{-1/2}.
\]
Both matrices are positive definite.
For a positive-definite matrix \(\bm B\),
write
\[
\|\bm x\|_{\bm B}
=
(\bm x^\top\bm B\bm x)^{1/2}.
\]
For \(\lambda>0\), define the
regularized effective dimension
\begin{equation}
\label{eq:general-effective-dimension}
d_n(\lambda)
=
\operatorname{tr}
\left\{
\bm R
(\bm R+\lambda\bm I_p)^{-1}
\right\},
\end{equation}
and the population ridge operator
\[
\bm T_\lambda
=
\left(
\bm\Sigma_{\bm Z\bm Z}
+
\lambda\bm D_{\bm Z}
\right)^{-1}
\bm\Sigma_{\bm Z\bm Z}.
\]

Recall the population quantities
\[
\bm a
=
\bm\Sigma_{\bm Z\bm Z}^{-1}
\bm\sigma_{\bm Z X},
\qquad
\bm v
=
\bm\Sigma_{\bm Z\bm Z}^{-1}\bm\xi,
\qquad
q
=
\bm\xi^\top
\bm\Sigma_{\bm Z\bm Z}^{-1}\bm\xi,
\]
and let
\[
\sigma_{YY\cdot(X,\bm Z)}
=
\sigma_{YY\cdot\bm Z}
-
\frac{\sigma_{YX\cdot\bm Z}^2}
{\sigma_{XX\cdot\bm Z}}.
\]
For the ridge parameter \(\lambda_n\),
define
\begin{align}
\varepsilon_{n_1}
&=
\left\{
\frac{
d_n(\lambda_n)+\log(p)
}{
n_1
}
\right\}^{1/2},
\label{eq:general-stochastic-rate}\\
b_{n_1}
&=
\max_{\bm u\in\{\bm a,\bm v\}}
\left\|
(\bm T_{\lambda_n}-\bm I_p)\bm u
\right\|_{\bm\Sigma_{\bm Z\bm Z}},
\label{eq:general-bias}\\
\eta_{n_1}
&=
b_{n_1}+\varepsilon_{n_1}.
\label{eq:general-direction-rate}
\end{align}
Here \(\varepsilon_{n_1}\) measures
the stochastic estimation scale,
whereas \(b_{n_1}\) measures
target-specific population ridge bias.
All asymptotic limits are taken as
\(n\to\infty\), with \(n_1\to\infty\);
both \(p=p_n\) and
\(\bm\Sigma=\bm\Sigma_n\) may vary
with \(n\).

\begin{assumption}[Basic regularity]
\label{assump:nondegeneracy}
There exist constants
\(0<c_0<C_0<\infty\), independent
of \(n\) and \(p\), such that
\[
c_0
\le
(\bm D_{\bm Z})_{jj}
\le C_0,
\qquad j=1,\ldots,p,
\]
\[
\min\left\{
\sigma_{XX\cdot\bm Z},
\sigma_{YY\cdot(X,\bm Z)},
q
\right\}
\ge c_0,
\]
and
\[
\max\{\sigma_{XX},\sigma_{YY},q\}
\le C_0.
\]
\end{assumption}

\begin{assumption}[Regularized complexity]
\label{assump:stochastic-complexity}
The regularized effective dimension
satisfies
\[
\frac{
d_n(\lambda_n)+\log(p)
}{
n_1
}
\longrightarrow0.
\]
\end{assumption}

\begin{assumption}[Population ridge approximation]
\label{assump:ridge-approximation}
The target-specific ridge bias satisfies
\[
b_{n_1}\longrightarrow0.
\]
\end{assumption}

\begin{assumption}[Rate condition for inference]
\label{assump:inference-rate}
For an independent evaluation sample
of size \(n_2\to\infty\),
\[
\sqrt{n_2}\,b_{n_1}^2
\longrightarrow0,
\qquad
\frac{\sqrt{n_2}}{n_1}
\{d_n(\lambda_n)+\log(p)\}
\longrightarrow0.
\]
\end{assumption}

Assumption~\ref{assump:nondegeneracy}
ensures nondegenerate residual
variances and stable normalization
of the population optimal direction,
while allowing the largest covariance
eigenvalue to increase with \(p\).
Assumptions~\ref{assump:stochastic-complexity}
and~\ref{assump:ridge-approximation}
control stochastic error and
population ridge bias, respectively,
and together imply \(\eta_{n_1}\to0\).
These conditions are related to
effective-dimension bounds in
random-design ridge regression
\citep{hsu2014random} and
approximation conditions in
regularized least-squares theory
\citep{caponnetto2007optimal}.

Assumption~\ref{assump:inference-rate}
strengthens these requirements
to the evaluation-sample inference
scale. Equivalently,
\[
\sqrt{n_2}\,\eta_{n_1}^2
\longrightarrow0,
\qquad\text{or}\qquad
\eta_{n_1}=o(n_2^{-1/4}).
\]
This quarter-rate requirement arises
from the second-order target-error
bound under rank-one alignment,
as in related orthogonal inference
arguments \citep{chernozhukov2018double}.
When \(n_1\asymp n_2\), it reduces to
\[
b_{n_1}=o(n_1^{-1/4}),
\qquad
d_n(\lambda_n)+\log(p)
=o(n_1^{1/2}).
\]

We first establish a quantitative
rate for estimating the population
optimal direction.

\begin{theorem}[Rate for the feasible ridge direction]
\label{thm:general-bridge}
Suppose that the Gaussian model
and Assumptions~\ref{assump:nondegeneracy}--%
\ref{assump:ridge-approximation} hold.
Let
\(\widehat{\bm P}_{\mathrm{ridge}}^\ast\)
be defined by \eqref{eq:P1_opt_Ri}.
Then, as \(n_1\to\infty\),
\[
\left\|
\widehat{\bm P}_{\mathrm{ridge}}^\ast
-
\bm P^\ast
\right\|_{\bm\Sigma_{\bm Z\bm Z}}
=
O_p(\eta_{n_1}),
\]
and
\[
0\le
Q(\bm P^\ast)
-
Q\!\left(
\widehat{\bm P}_{\mathrm{ridge}}^\ast
\right)
=
O_p(\eta_{n_1}^2).
\]
\end{theorem}

Theorem~\ref{thm:general-bridge}
provides a direction-estimation
rate and a quadratic bound on
the loss of population retention.
Under rank-one alignment, this
retention loss also controls the
error in the original inferential
target.

\begin{proposition}[Second-order control of the target error]
\label{prop:second-order-target-error}
Suppose that the Gaussian model
and Assumption~\ref{assump:nondegeneracy}
hold. Let
\(\bm P^\ast=\bm v/\sqrt q\)
satisfy the rank-one alignment condition
\[
\bm a\in\operatorname{col}(\bm P^\ast).
\]
Then, for every nonzero
\(\bm P\in\mathbb R^p\),
\begin{align*}
|\rho(\bm P)-\rho|
&\le
\frac{2C_0}{c_0q}
\left\{
Q(\bm P^\ast)-Q(\bm P)
\right\}\\
&\le
\frac{2C_0}{c_0}
\|\bm P-\bm P^\ast\|
_{\bm\Sigma_{\bm Z\bm Z}}^2.
\end{align*}
\end{proposition}

Combining Theorem~\ref{thm:general-bridge}
and Proposition~%
\ref{prop:second-order-target-error}
gives, under their joint conditions,
\[
\left|
\rho\!\left(
\widehat{\bm P}_{\mathrm{ridge}}^\ast
\right)-\rho
\right|
=
O_p(\eta_{n_1}^2).
\]
Under
Assumption~\ref{assump:inference-rate},
this further implies
\[
\sqrt{n_2}
\left\{
\rho\!\left(
\widehat{\bm P}_{\mathrm{ridge}}^\ast
\right)-\rho
\right\}
=o_p(1).
\]
Thus the projection-induced target
error is negligible at the
evaluation-sample inference scale.

\paragraph{An explicit coordinate-dense model.}

We verify the learning conditions
in a Gaussian model with
\(p\gg n\) and coordinate-dense
regression coefficients.
Let \(\bm\ell\in\mathbb R^p\)
satisfy
\[
0<c_\ell\le|\ell_j|\le C_\ell<\infty,
\qquad j=1,\ldots,p,
\]
uniformly in \(n\) and \(p\), and suppose
\[
\bm Z
\sim
N_p(\bm 0,\bm I_p+\bm\ell\bm\ell^\top).
\]
Define
\[
\mu=1+\|\bm\ell\|_2^2,
\qquad
\bm e=\frac{\bm\ell}{\|\bm\ell\|_2},
\qquad
H=\frac{\bm e^\top\bm Z}{\sqrt\mu},
\]
so that \(\operatorname{var}(H)=1\).
Let
\[
X=H+\varepsilon_2,
\qquad
Y=-H+\varepsilon_1,
\]
where
\[
\begin{pmatrix}
\varepsilon_1\\
\varepsilon_2
\end{pmatrix}
\sim
N_2\left(
\bm 0,
\begin{pmatrix}
1 & \rho_0\\
\rho_0 & 1
\end{pmatrix}
\right),
\qquad
|\rho_0|\le\rho_{\max}<1,
\]
independently of \(\bm Z\).

The population regression
coefficients of \(X\) and \(Y\)
on \(\bm Z\) are
\(\bm a=\bm e/\sqrt\mu\)
and \(-\bm a\), respectively.
Both are coordinate-dense, and
\(\|\bm a\|_{\bm\Sigma_{\bm Z\bm Z}}=1\).
Moreover,
\[
\rho=\rho_0,
\qquad
\bm v=-(1+\rho_0)\bm a,
\qquad
q=(1+\rho_0)^2,
\qquad
\bm P^\ast=-\bm a.
\]
Thus rank-one alignment holds.
The identities
\[
\sigma_{XX\cdot\bm Z}=1,
\qquad
\sigma_{YY\cdot(X,\bm Z)}=1-\rho_0^2,
\qquad
\sigma_{XX}=\sigma_{YY}=2,
\]
together with
\[
(\bm D_{\bm Z})_{jj}=1+\ell_j^2,
\qquad
q\ge(1-\rho_{\max})^2,
\]
verify
Assumption~\ref{assump:nondegeneracy}.

To establish the learning rates,
set
\[
\bm w
=
\bm D_{\bm Z}^{-1/2}\bm\ell.
\]
The standardized covariance matrix is
\[
\bm R
=
\bm D_{\bm Z}^{-1}
+
\bm w\bm w^\top.
\]
Let its ordered eigenvalues be
\(\nu_1\ge\cdots\ge\nu_p>0\).
The loading bounds and rank-one
structure imply
\[
\nu_1\asymp p,
\qquad
\max_{2\le j\le p}\nu_j\le C.
\]
Consequently,
\[
d_n(\lambda)
=
\sum_{j=1}^p
\frac{\nu_j}{\nu_j+\lambda}
\le
1+\frac{Cp}{\lambda}.
\]

For \(\bm u\in\{\bm a,\bm v\}\),
\[
\|\bm u\|_{\bm\Sigma_{\bm Z\bm Z}}\le C,
\qquad
\|\bm D_{\bm Z}^{1/2}\bm u\|_2
\le Cp^{-1/2}.
\]
Let
\[
\bm R
=
\bm U
\operatorname{diag}(\nu_1,\ldots,\nu_p)
\bm U^\top,
\qquad
\bm\theta_{\bm u}
=
\bm U^\top
\bm D_{\bm Z}^{1/2}\bm u.
\]
The spectral representation of
the population ridge bias gives
\[
\begin{aligned}
&
\|(\bm T_\lambda-\bm I_p)\bm u
\|_{\bm\Sigma_{\bm Z\bm Z}}^2\\
&\quad=
\sum_{j=1}^p
\nu_j
\left(
\frac{\lambda}{\nu_j+\lambda}
\right)^2
\theta_{\bm u,j}^2\\
&\quad\le
C\left\{
\left(\frac{\lambda}{p}\right)^2
+\frac1p
\right\}.
\end{aligned}
\]
Here the leading term uses
\(\nu_1\asymp p\) and
\(\nu_1\theta_{\bm u,1}^2\le C\);
the remaining terms use
\(\max_{j\ge2}\nu_j\le C\) and
\(\sum_{j=2}^p\theta_{\bm u,j}^2
\le Cp^{-1}\).
Hence
\[
b_{n_1}
\le
C\left(
\frac{\lambda_n}{p}
+
p^{-1/2}
\right).
\]

For a concrete growth regime,
choose
\[
n_1=n_2=N,
\qquad
p=N^2,
\qquad
\lambda_n=pN^{-1/3}.
\]
The preceding bounds give
\[
d_n(\lambda_n)=O(N^{1/3}),
\qquad
b_N=O(N^{-1/3}),
\qquad
\eta_N=O(N^{-1/3}),
\]
so that
\[
\sqrt N\,\eta_N^2
=
O(N^{-1/6})
\longrightarrow0.
\]
Thus
Assumptions~\ref{assump:nondegeneracy}--%
\ref{assump:inference-rate}
hold despite \(p\gg n_1+n_2\).
Theorem~\ref{thm:general-bridge}
and Proposition~%
\ref{prop:second-order-target-error}
therefore yield
\[
\left\|
\widehat{\bm P}_{\mathrm{ridge}}^\ast
-
\bm P^\ast
\right\|_{\bm\Sigma_{\bm Z\bm Z}}
=
O_p(N^{-1/3}),
\]
and
\[
\left|
\rho(\widehat{\bm P}_{\mathrm{ridge}}^\ast)
-\rho
\right|
=
O_p(N^{-2/3})
=
o_p(N^{-1/2}).
\]
This example demonstrates that
the inference-scale conditions
can hold without coordinate sparsity
when the target-relevant directions
have sufficiently learnable
population structure.

\paragraph{Scope of the learning conditions.}

The learning conditions do not cover
unstructured isotropic models with
\(p\ge n_1\) and \(q\) bounded away
from zero. For example, if
\(\bm\Sigma_{\bm Z\bm Z}=\bm I_p\)
and \(\bm a=\bm 0\), then
\[
b_{n_1}
=
\frac{\lambda_n}{1+\lambda_n}
\sqrt q,
\qquad
d_n(\lambda_n)
=
\frac{p}{1+\lambda_n}.
\]
Since \(q\ge c_0\),
the requirement \(b_{n_1}\to0\)
forces \(\lambda_n\to0\).
When \(p\ge n_1\), this prevents
\(d_n(\lambda_n)/n_1\) from vanishing.
Such models therefore lie outside
the sufficient conditions of
Theorem~\ref{thm:general-bridge}.

In contrast, the coordinate-dense
model above satisfies
\[
\lambda_{\max}
(\bm\Sigma_{\bm Z\bm Z})
\asymp p,
\]
placing it outside the
bounded-spectrum parameter class
considered by
\citet{ren2015asymptotic}.
Our results do not claim uniform
inference over unrestricted dense
covariance models or contradict
lower bounds established for
different parameter classes.
Rather, they establish inference
under explicit conditions ensuring
that the relevant population
directions are learnable.

We next establish feasible inference
for the original partial correlation.
For any nonzero direction \(\bm P\),
let \(r_{\mathcal I_2}(\bm P)\)
denote the sample partial correlation
computed on the evaluation sample,
including an intercept. Write
\[
\widehat{\bm P}
=
\widehat{\bm P}_{\mathrm{ridge}}^\ast,
\qquad
\widehat\rho
=
r_{\mathcal I_2}(\widehat{\bm P}),
\qquad
\widetilde\rho
=
r_{\mathcal I_2}(\bm P^\ast).
\]
Both estimators use the same
evaluation sample.
Let \(\mathcal F_1\) be the
sigma-field generated by the
sample partition and training
observations.

\begin{corollary}[Feasible OPE inference]
\label{cor:feasible-ope}
Suppose that the Gaussian model
and Assumptions~\ref{assump:nondegeneracy}--%
\ref{assump:inference-rate} hold,
and that
\[
\bm a\in\operatorname{col}(\bm P^\ast).
\]
Then
\[
\sqrt{n_2}
(\widehat\rho-\widetilde\rho)
=o_p(1),
\]
and
\[
\frac{
\sqrt{n_2}(\widehat\rho-\rho)
}{
1-\widehat\rho^{\,2}
}
\xrightarrow{d}N(0,1).
\]
\end{corollary}

Corollary~\ref{cor:feasible-ope}
establishes oracle equivalence
and asymptotic normality of
the feasible OPE.
An asymptotic standard error is
\[
\widehat{\operatorname{se}}
=
\frac{1-\widehat\rho^{\,2}}{\sqrt{n_2}}.
\]
For fixed \(\alpha\in(0,1)\),
define the confidence interval
\begin{equation}
\label{eq:feasible-ope-ci}
\mathcal C_{1-\alpha}
=
\left[
\widehat\rho
-
z_{1-\alpha/2}
\widehat{\operatorname{se}},
\;
\widehat\rho
+
z_{1-\alpha/2}
\widehat{\operatorname{se}}
\right]
\cap[-1,1],
\end{equation}
where \(z_\tau\) denotes the
\(\tau\)-quantile of the standard
normal distribution.
Under the conditions of
Corollary~\ref{cor:feasible-ope},
\[
\Pr\{\rho\in\mathcal C_{1-\alpha}\}
=
1-\alpha+o(1).
\]

We now consider testing the
original null \(H_0:\rho=0\).
Conditional on \(\mathcal F_1\),
the learned direction is fixed.
Under \(H_0\), the second-order
target-error bound and
Assumption~\ref{assump:inference-rate}
give
\[
\sqrt{n_2}\,
\rho(\widehat{\bm P})
=o_p(1).
\]
Together with the conditional
noncentral \(F\)-representation,
this yields asymptotic validity
for the original null.

\begin{corollary}[Testing the original null]
\label{cor:original-null-test}
Suppose that the conditions of
Corollary~\ref{cor:feasible-ope}
hold. For fixed
\(\alpha\in(0,1)\), define
\[
F_{\mathrm{feas}}
=
(n_2-3)
\frac{\widehat\rho^{\,2}}
{1-\widehat\rho^{\,2}},
\]
and let \(c_{\alpha,n_2}\)
be the \((1-\alpha)\)-quantile
of the central
\(F_{1,n_2-3}\) distribution.
Under \(H_0:\rho=0\),
\[
\Pr\left\{
F_{\mathrm{feas}}>c_{\alpha,n_2}
\mid\mathcal F_1
\right\}
=
\alpha+o_p(1).
\]
Consequently,
\[
\Pr\left\{
F_{\mathrm{feas}}>c_{\alpha,n_2}
\right\}
=
\alpha+o(1).
\]
\end{corollary}

The original-null guarantee is
asymptotic. Exact conditional
central \(F\) calibration holds
under the projected null
\(\rho(\widehat{\bm P})=0\).

Algorithm~\ref{alg:feasible-ope}
summarizes the split-sample procedure.
Let \(G_{1,\nu}\) denote the
distribution function of the
central \(F_{1,\nu}\) distribution.

\begin{algorithm}[tbp]
\caption{Split-sample inference for partial correlation}
\label{alg:feasible-ope}
\begin{algorithmic}[1]

\Require
Observations \(\{\bm W_i\}_{i=1}^n\);
training size \(n_1\ge3\), with
\(n_2=n-n_1>3\);
deterministic ridge parameter
\(\lambda_n>0\);
significance level \(\alpha\in(0,1)\).

\Ensure
Estimate \(\widehat\rho\),
confidence interval
\(\mathcal C_{1-\alpha}\),
and test \(p\)-value
\(p_{\mathrm{val}}\).

\State Partition the observations
into \(\mathcal I_1\) and
\(\mathcal I_2\), independently
of their values, with
\(|\mathcal I_k|=n_k\).

\State Using \(\mathcal I_1\),
construct
\(\widehat{\bm P}
=\widehat{\bm P}_{\mathrm{ridge}}^\ast\)
according to
\eqref{eq:P1_opt_Ri}, using
centered sample covariances
normalized by \(1/(n_1-1)\).

\State For \(i\in\mathcal I_2\),
compute
\(U_i=\widehat{\bm P}^\top\bm Z_i\).

\State Using \(\mathcal I_2\),
compute
\(\widehat\rho
=r_{\mathcal I_2}(\widehat{\bm P})\),
including an intercept.

\State Set
\[
\widehat{\operatorname{se}}
=
\frac{1-\widehat\rho^{\,2}}
{\sqrt{n_2}},
\]
and construct
\(\mathcal C_{1-\alpha}\)
using \eqref{eq:feasible-ope-ci}.

\State Compute
\[
F_{\mathrm{feas}}
=
(n_2-3)
\frac{\widehat\rho^{\,2}}
{1-\widehat\rho^{\,2}},
\qquad
p_{\mathrm{val}}
=
1-G_{1,n_2-3}(F_{\mathrm{feas}}).
\]

\State Reject \(H_0:\rho=0\)
if \(p_{\mathrm{val}}\le\alpha\).

\State \Return
\(\widehat\rho\),
\(\mathcal C_{1-\alpha}\),
and \(p_{\mathrm{val}}\).

\end{algorithmic}
\end{algorithm}

For efficient computation when
\(p>n_1\), the ridge products
\(\widehat{\bm M}
\widehat{\bm\sigma}_{\bm Z X}\)
and
\(\widehat{\bm M}
\widehat{\bm\sigma}_{\bm Z Y}\)
can be obtained through an
\(n_1\times n_1\) linear system.
Let
\(\bm Z_{1,c}\in\mathbb R^{n_1\times p}\)
be the centered training covariate
matrix and define
\[
\widetilde{\bm Z}_1
=
\bm Z_{1,c}
\widehat{\bm D}_{\bm Z}^{-1/2},
\qquad
\bm b_B
=
(B_i-\overline B_1)_{i\in\mathcal I_1},
\quad B\in\{X,Y\},
\]
where the observations are
ordered consistently with the
rows of \(\bm Z_{1,c}\).

The ridge identity gives
\begin{equation}
\label{eq:ridge-dual-implementation}
\begin{aligned}
&\widehat{\bm M}
\widehat{\bm\sigma}_{\bm Z B}\\
&\quad=
\widehat{\bm D}_{\bm Z}^{-1/2}
\widetilde{\bm Z}_1^\top
\left\{
\widetilde{\bm Z}_1
\widetilde{\bm Z}_1^\top
+
(n_1-1)\lambda_n\bm I_{n_1}
\right\}^{-1}
\bm b_B.
\end{aligned}
\end{equation}
Since \(\lambda_n>0\),
the matrix in braces is
positive definite, and a
single factorization suffices
for both responses.

The ridge products determine
\[
\begin{aligned}
\widehat{\bm M}\widehat{\bm\xi}
&=
\widehat{\bm M}
\widehat{\bm\sigma}_{\bm Z Y}
-
\widehat\gamma\,
\widehat{\bm M}
\widehat{\bm\sigma}_{\bm Z X},\\
\widehat q
&=
\widehat{\bm\xi}^\top
\bigl(
\widehat{\bm M}\widehat{\bm\xi}
\bigr).
\end{aligned}
\]
Thus
\eqref{eq:ridge-dual-implementation}
provides an algebraically
equivalent computation of
the feasible direction in
\eqref{eq:P1_opt_Ri},
without forming the full
\(p\times p\) covariance matrix.

Section~\ref{s4} examines
the finite-sample performance
of the resulting estimator
and inferential procedures.

\section{Numerical experiments}\label{s4}
We examine the finite-sample behavior of the
feasible reduction in three related Gaussian
designs.
Design 1 provides a dense single-factor
baseline.
Design 2 adds a stronger factor unrelated
to the target, while Design 3 restricts
the factor loadings and regression
coefficients to a common sparse support.

Projection diagnostics in Designs 1--2
assess direction learning, target preservation,
and agreement with oracle inference.
Comparisons with existing methods assess
estimation and testing in all three designs.
The response variances and marginal
correlations are matched across designs,
allowing us to examine how covariate
structure and regression support affect
performance under a common association model.

\subsection{Gaussian designs and experimental objectives}

Throughout, we fix the factor-loading
magnitude at \(\tau=1\).
Independent observations satisfy
\begin{equation}\label{eq:sim-responses}
 X_i=\sqrt{\delta}\,H_i+\varepsilon_{i2},
 \qquad
 Y_i=-\sqrt{\delta}\,H_i+\varepsilon_{i1},
 \qquad
 \operatorname{var}(H_i)=1,
 \qquad
 \delta=0.2,
\end{equation}
where \(H_i\) is a linear function of the
Gaussian conditioning vector \(\bm{Z}_i\).
The errors satisfy
\[
 \begin{pmatrix}
  \varepsilon_{i1}\\
  \varepsilon_{i2}
 \end{pmatrix}
 \sim
 N_2\left(
  \bm{0},
  \begin{pmatrix}
   1 & \rho_0\\
   \rho_0 & 1
  \end{pmatrix}
 \right),
 \qquad
 \rho_0\in(-1,1),
\]
independently of \(\bm{Z}_i\).
The covariate contribution to each response
therefore has variance \(\delta\), and the
original population partial correlation
is \(\rho=\rho_0\).

In all three designs,
\[
 \bm{\beta}_Y=-\bm{\beta}_X,
 \qquad
 \bm{\beta}_X^{\top}
 \bm{\Sigma}_{\bm{Z}\bm{Z}}
 \bm{\beta}_X=\delta.
\]
Thus rank-one alignment holds, with
\[
 \bm{v}=-(1+\rho_0)\bm{\beta}_X,
 \qquad
 q=\delta(1+\rho_0)^2,
 \qquad
 \bm{P}^{\ast}
 =
 -\frac{\bm{\beta}_X}{\sqrt{\delta}}.
\]
The oracle direction has unit prediction
norm and preserves the original target.
This population property is distinct
from the accuracy with which its
direction can be learned from finite
training samples.

The marginal correlation is
\begin{equation}\label{eq:sim-marginal-correlation}
 \operatorname{Corr}(Y_i,X_i)
 =
 \frac{\rho_0-\delta}{1+\delta}.
\end{equation}
For \(\rho_0\in\{0,-0.2,0.2\}\), the
corresponding marginal correlations are
\(-1/6\), \(-1/3\), and \(0\), respectively.
These scenarios assess null calibration
in the presence of covariate-induced
association and the detection of
conditional association with and
without marginal cancellation.

\paragraph{Design 1: A dense single-factor baseline.}

Generate
\begin{equation}\label{eq:sim-design1-factor-model}
 \bm{Z}_i
 =
 \bm{\ell}F_i+\bm{\zeta}_i,
 \qquad
 F_i\sim N(0,1),
 \qquad
 \bm{\zeta}_i\sim N_p(\bm{0},\bm{I}_p),
\end{equation}
with independent \(F_i\) and
\(\bm{\zeta}_i\).
With \(\tau=1\), the fixed loadings are
\[
 \ell_j=\tau(-1)^j=(-1)^j,
 \qquad j=1,\ldots,p.
\]
The covariate covariance matrix is
\[
 \bm{\Sigma}_{\bm{Z}\bm{Z}}
 =
 \bm{I}_p+\bm{\ell}\bm{\ell}^{\top}.
\]
Define
\begin{equation}\label{eq:sim-design1-index}
 \mu=1+\|\bm{\ell}\|_2^2=1+p,
 \qquad
 \bm{e}=\frac{\bm{\ell}}{\sqrt{p}},
 \qquad
 H_i
 =
 \frac{\bm{e}^{\top}\bm{Z}_i}{\sqrt{\mu}}.
\end{equation}
Then
\[
 \bm{\beta}_X
 =
 \frac{\sqrt{\delta}\,\bm{e}}{\sqrt{\mu}},
 \qquad
 \bm{\beta}_Y=-\bm{\beta}_X.
\]
Both regression vectors are coordinate-dense.
The leading covariance eigenvalue is
\(1+p\), and its eigenvector spans the
target-relevant regression direction.

This design provides a baseline in
which the common regression component
coincides with the only strong covariance
direction.
We compare Ridge with Oracle using
prediction-norm direction error,
standardized population target error,
and the standardized difference between
their estimates on the same evaluation
sample.

\paragraph{Design 2: A stronger irrelevant factor.}

For even \(p\), define the orthonormal
vectors
\[
 \bm{u}_1=\frac{\bm{1}_p}{\sqrt{p}},
 \qquad
 \bm{u}_2
 =
 \frac{1}{\sqrt{p}}
 \bigl((-1)^j\bigr)_{j=1}^{p}.
\]
Using the same loading magnitude
\(\tau=1\), generate
\[
 \bm{Z}_i
 =
 \sqrt{2p}\,\bm{u}_1F_{i1}
 +
 \sqrt{p}\,\bm{u}_2F_{i2}
 +
 \bm{\zeta}_i,
\]
where \(F_{i1},F_{i2}\sim N(0,1)\) and
\(\bm{\zeta}_i\sim N_p(\bm{0},\bm{I}_p)\)
are mutually independent.
Hence
\[
 \bm{\Sigma}_{\bm{Z}\bm{Z}}
 =
 \bm{I}_p
 +
 2p\,\bm{u}_1\bm{u}_1^{\top}
 +
 p\,\bm{u}_2\bm{u}_2^{\top}.
\]
Set
\[
 H_i
 =
 \frac{\bm{u}_2^{\top}\bm{Z}_i}
      {\sqrt{1+p}},
 \qquad
 \bm{\beta}_X
 =
 \frac{\sqrt{\delta}\,\bm{u}_2}
      {\sqrt{1+p}}.
\]

The target-relevant eigenvalue \(1+p\)
matches that of Design 1.
The additional direction \(\bm{u}_1\)
has the larger eigenvalue \(1+2p\),
but its score is independent of \(H_i\)
and of the responses.
This construction examines whether the
learned reduction retains the relevant
component despite stronger irrelevant
variation.

In addition to the Ridge--Oracle
diagnostics, we compare target and
irrelevant-direction retention for
Oracle, Ridge, PCA1, and PCA2.
PCA1 uses the leading eigenvector of
the centered training-sample covariance
matrix; PCA2 uses its first two
eigenvectors jointly.
Oracle, Ridge, and PCA1 are
one-dimensional, whereas PCA2 is
two-dimensional.
These benchmarks distinguish
target-specific reduction from adjustment
for the two-factor space.

\paragraph{Design 3: A sparse counterpart.}

For each dimension \(p\), sample a
support \(S\) uniformly from the subsets
of \(\{1,\ldots,p\}\) of size \(s_0\)
and retain it throughout the experiment.
Use the single-factor construction
in \eqref{eq:sim-design1-factor-model}
with
\[
 \ell_j
 =
 (-1)^j\mathbf{1}\{j\in S\}.
\]
Define
\[
 \mu_s=1+\|\bm{\ell}\|_2^2=1+s_0,
 \qquad
 \bm{e}_s
 =
 \frac{\bm{\ell}}{\sqrt{s_0}},
 \qquad
 H_i
 =
 \frac{\bm{e}_s^{\top}\bm{Z}_i}
      {\sqrt{\mu_s}}.
\]
The regression coefficients are
\[
 \bm{\beta}_X
 =
 \frac{\sqrt{\delta}\,\bm{e}_s}
      {\sqrt{\mu_s}},
 \qquad
 \bm{\beta}_Y=-\bm{\beta}_X.
\]
Both have exactly \(s_0\) nonzero
coordinates on the same support.

For fixed \(s_0\), this design has sparse
nuisance regressions and a sparse precision
matrix, while retaining rank-one alignment
and a target-preserving Oracle.
Restricting the support reduces the
relevant eigenvalue from \(1+p\) to
\(1+s_0\), so the comparison varies both
sparsity and factor strength.
Under \(\lambda_n=p\,n_1^{-1/3}\), the
condition \(b_{n_1}\to0\) fails along
sequences with \(p\geq n_1\to\infty\)
and fixed \(s_0\).
Design 3 therefore examines empirical
performance beyond the established
sufficient learning conditions, using
Oracle as the target-preserving benchmark.

\paragraph{Experimental settings.}

For estimation and testing in Designs 1--3,
we use the full factorial grid
\[
 n\in\{500,1000\},
 \qquad
 p\in\{500,1000\},
 \qquad
 \rho_0\in\{0,-0.2,0.2\},
\]
giving 12 configurations per design.
Here \(n\) is the total sample size and
\(p\) is the conditioning dimension.
We set \(s_0=30\) in Design 3.
For each design and dimension, the
covariate distribution and regression
coefficients, including the sampled
support in Design 3, are held fixed
across sample sizes, replications,
target correlations, and tuning settings.
We vary \(\rho_0\) solely through the
correlation between the regression errors.

Each configuration uses a common sequence
of \(R=500\) independent Monte Carlo
datasets, with replications distributed
across eight parallel workers.
Estimation comparisons include OPE,
Oracle, SPACE, scaled Lasso, and RSZZ;
testing comparisons include OPT,
Oracle, P-DCov, and RSZZ.
To limit computational cost, SPACE is
evaluated on the first
\(R_{\mathrm{SPACE}}=50\) datasets
in each configuration.
All other methods use all 500 datasets.
Thus, methods evaluated in the same
replication receive the same generated
data, and the SPACE subset is determined
by replication number rather than
observed performance.

Following the preliminary study in
Section~\ref{sec:tuning}, we fix the
training fraction and ridge multiplier
across the subsequent experiments:
\[
 n_1=\lfloor0.2n\rfloor,
 \qquad
 n_2=n-n_1,
 \qquad
 \lambda_n=c\,p\,n_1^{-1/3},
 \qquad
 c=1.
\]
The random partition is independent
of the observed values.
OPE and OPT learn their direction
exclusively from the training sample
and perform estimation or testing
on the independent evaluation sample.
Oracle uses the target-preserving
population direction and exactly the
same evaluation observations.
The competing methods use all \(n\)
observations with their respective
fitting, tuning, and calibration
procedures.
P-DCov uses \(B=199\) residual
permutations per test.

The training sample sizes are 100
and 200 when \(n=500\) and \(n=1000\),
respectively.
Thus \(p>n_1\) throughout the main grid.
Since a centered training covariance
matrix has rank at most \(n_1-1\),
it is singular in every configuration.

For the projection diagnostics in
Designs 1--2, we fix \(p=2000\)
and \(\rho_0=0\), and vary
\[
 n\in\{200,500,1000,2000\}.
\]
Using the same split and ridge rules
gives training sample sizes
\[
 n_1\in\{40,100,200,400\}.
\]
Each configuration uses \(R=500\)
replications.
The Design 2 retention comparison
uses \(n=p=2000\) and \(\rho_0=0\).
All sample principal components are
estimated using only the training
covariates.

We additionally assess testing in
Design 1 with \(n=500\), \(p=10000\),
the same three values of \(\rho_0\),
and \(R=500\) replications per
configuration.
This extension retains \(\tau=1\),
\(\delta=0.2\), and the same split
and ridge rules, and examines
performance when \(p\gg n\).
To limit computational cost, we compare
OPT, Oracle, P-DCov, and RSZZ without
extending the full point-estimation
comparison to this dimension.

All tests use nominal level
\(\alpha=0.05\).
We report Monte Carlo mean estimates
and empirical rejection probabilities.
Null rejection probabilities assess
type-I error, while alternative
rejection probabilities assess power
and are interpreted alongside null
calibration.
Computation times are recorded both
as mean elapsed times per replication
and as cumulative elapsed times over
the replications used by each method.
When replication counts differ,
computational comparisons use the
mean time per replication.
Oracle timing excludes direction
learning, and parallel wall-clock
time is recorded separately.

\subsection{Evaluation metrics}
\label{sec:simulation-metrics}

\paragraph{Projection diagnostics.}

In Designs 1--2, let
\(\widehat{\bm{P}}_b\) denote the feasible
ridge direction learned in replication
\(b=1,\ldots,R\).
The feasible and oracle estimates,
\[
 \widehat{\rho}_b
 =
 r_{\mathcal{I}_{2,b}}(\widehat{\bm{P}}_b),
 \qquad
 \widetilde{\rho}_b
 =
 r_{\mathcal{I}_{2,b}}(\bm{P}^{\ast}),
\]
use the same evaluation sample of size
\(n_2\), so their difference isolates
the effect of estimating the projection.

Using the oracle asymptotic standard error
\[
 s_{\mathrm{or}}
 =
 \frac{1-\rho_0^2}{\sqrt{n_2}},
\]
we define
\begin{equation}\label{eq:simulation-diagnostics}
\begin{aligned}
 d_b
 &=
 \left\|
  \widehat{\bm{P}}_b-\bm{P}^{\ast}
 \right\|_{\bm{\Sigma}_{\bm{Z}\bm{Z}}},
 \\
 t_b
 &=
 \frac{
  |\rho(\widehat{\bm{P}}_b)-\rho_0|
 }{s_{\mathrm{or}}},
 \\
 g_b
 &=
 \frac{
  \widehat{\rho}_b-\widetilde{\rho}_b
 }{s_{\mathrm{or}}}.
\end{aligned}
\end{equation}
These quantities measure prediction-norm
direction error, standardized population
target error, and the standardized
feasible-to-oracle difference, respectively.
Both directions retain the signs and
normalizations specified in their
definitions.
Since
\(\|\bm{P}^{\ast}\|_{\bm{\Sigma}_{\bm{Z}\bm{Z}}}=1\),
the oracle projection score provides
a unit-variance reference for \(d_b\).

We summarize the diagnostics by
\begin{equation}\label{eq:sim-diagnostic-summaries}
 \bar d
 =
 \frac1R\sum_{b=1}^{R}d_b,
 \qquad
 \bar t
 =
 \frac1R\sum_{b=1}^{R}t_b,
 \qquad
 g_{\mathrm{RMS}}
 =
 \left(
  \frac1R\sum_{b=1}^{R}g_b^2
 \right)^{1/2}.
\end{equation}
The root mean square avoids cancellation
of signed oracle discrepancies.
Values of \(\bar t\) and
\(g_{\mathrm{RMS}}\) well below one
indicate errors small relative to
the oracle sampling scale.

In Design 2, we also measure retention
of the target index and the irrelevant
factor score.
For a realized full-column-rank
projection \(\bm{P}\), define
\begin{equation}\label{eq:sim-retention-metrics}
\begin{aligned}
 \mathcal{R}_{\mathrm{target}}(\bm{P})
 &=
 1-
 \operatorname{var}
 (H_i\mid\bm{P}^{\top}\bm{Z}_i)
 =
 \frac{Q(\bm{P})}{q},
 \\
 \mathcal{R}_{\mathrm{irrelevant}}(\bm{P})
 &=
 1-
 \frac{
  \operatorname{var}
  (\bm{u}_1^{\top}\bm{Z}_i
   \mid\bm{P}^{\top}\bm{Z}_i)
 }{1+2p}.
\end{aligned}
\end{equation}
These population quantities are evaluated
with the realized projection held fixed.
Because
\(\operatorname{var}(H_i)=1\) and
\(\operatorname{var}(\bm{u}_1^{\top}\bm{Z}_i)=1+2p\),
both ratios lie in \([0,1]\) and represent
fractions of component variance retained
by the projected covariates.

We report their Monte Carlo averages
for Oracle, Ridge, PCA1, and PCA2,
alongside the projection dimension.
High target retention and low
irrelevant-direction retention characterize
a target-specific one-dimensional
reduction.
PCA2 can retain both components through
its two-dimensional construction.
The population covariance matrices and
oracle directions used in these diagnostics
are reserved for simulation evaluation
and are not supplied to feasible fitting
or tuning.

\paragraph{Testing summaries.}

All testing methods use \(R=500\)
replications.
For method \(k\), let \(p_{b,k}\)
denote its \(p\)-value in replication \(b\).
We report the empirical rejection
probability
\begin{equation}\label{eq:sim-rejection-probability}
 \widehat{\pi}_k(\rho_0;\alpha)
 =
 \frac1R
 \sum_{b=1}^{R}
 \mathbf{1}\{p_{b,k}\leq\alpha\}.
\end{equation}
Under \(\rho_0=0\), this estimates
type-I error and is compared with
the nominal level \(\alpha\).
Under \(\rho_0\ne0\), it estimates
power, which is interpreted alongside
the corresponding null rejection
probability.

\subsection{Choice of the training fraction and ridge multiplier}
\label{sec:tuning}

Training allocation balances direction learning
against evaluation-sample size, while ridge
regularization balances sampling variability
against bias.
We examine these choices under Design 1 with
\(n=p=500\), \(\tau=1\), and \(\delta=0.2\),
using \(R_{\mathrm{sel}}=500\) replications
per \(\rho_0\in\{0,-0.2,0.2\}\).
We consider
\[
 \kappa\in\{0.2,0.3,\ldots,0.8\},
 \qquad
 c\in\{0.5,1,2\},
\]
with
\[
 n_1=\lfloor\kappa n\rfloor,
 \qquad
 n_2=n-n_1,
 \qquad
 \lambda_n=c\,p\,n_1^{-1/3}.
\]
Candidates share the same simulated datasets.
At nominal level \(\alpha=0.05\), we retain
candidates for which all fits succeed and
the empirical null rejection rate does not
exceed \(0.05\), then maximize the smaller
empirical power at \(\rho_0=\pm0.2\).

\begin{figure}[t]
 \centering
 \includegraphics[width=\linewidth]
 {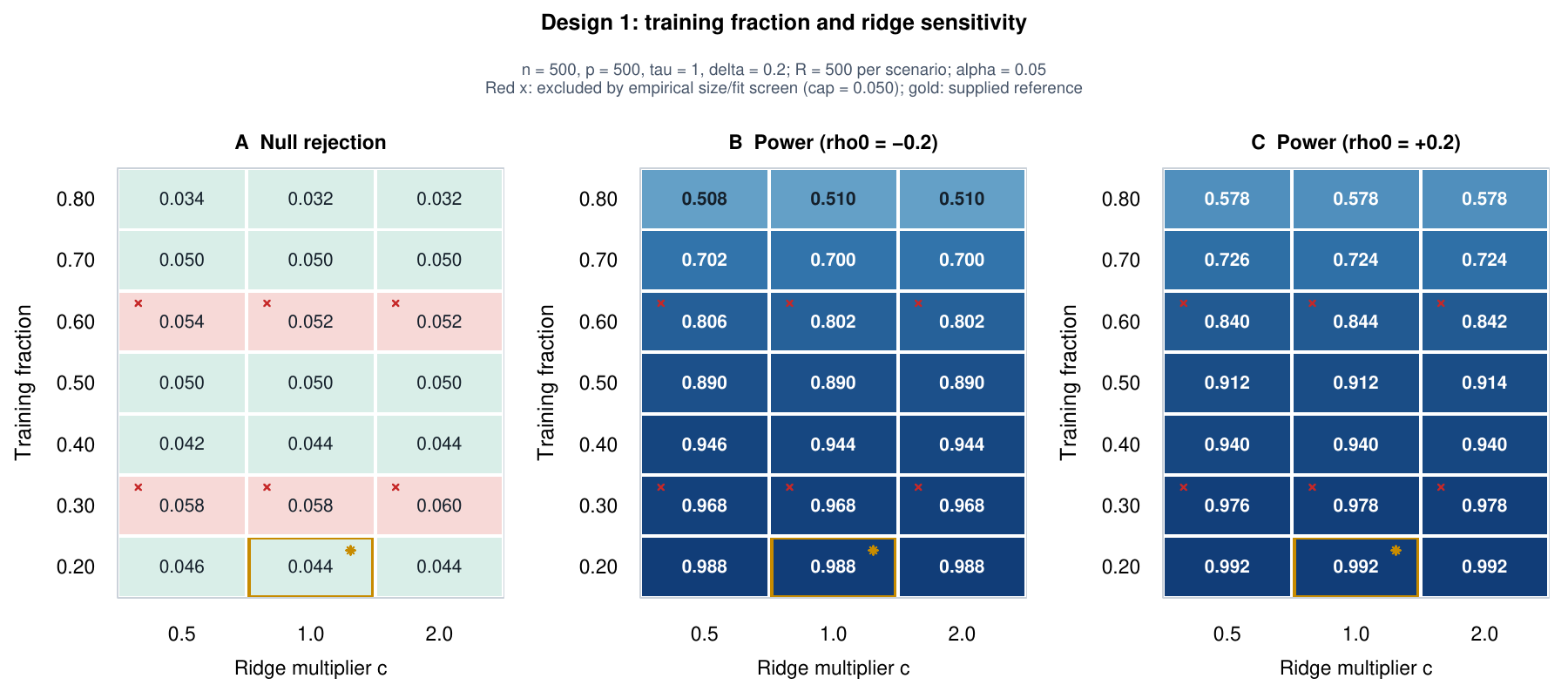}
 \caption{
 Preliminary split and ridge selection
 for OPT in Design 1 with \(n=p=500\),
 \(\tau=1\), and \(\delta=0.2\).
 Each candidate uses
 \(R_{\mathrm{sel}}=500\) replications
 per scenario at nominal level
 \(\alpha=0.05\).
 Panels A--C report empirical null
 rejection rates and power at
 \(\rho_0=-0.2\) and \(\rho_0=0.2\),
 respectively.
 Red crosses mark excluded candidates;
 gold borders mark the adopted setting
 \((\kappa,c)=(0.2,1)\).
 }
 \label{fig:split-ridge-selection}
\end{figure}

Figure~\ref{fig:split-ridge-selection}
shows that power decreases as the training
fraction increases, consistent with fewer
observations remaining for evaluation,
whereas the ridge multiplier has little
effect over the examined grid.
At \(\kappa=0.2\), all three multipliers
pass the empirical screen and attain powers
of \(0.988\) and \(0.992\) under the negative
and positive alternatives, respectively.
We therefore adopt \(\kappa=0.2\) and retain
\(c=1\) as the reference multiplier among
these equal-power candidates.
This gives \(n_1=100\), \(n_2=400\), and an
empirical null rejection rate of \(0.044\).
These values are fixed before the main
experiments, where projection learning uses
the training sample and inference uses only
the independent evaluation sample.

\subsection{Simulation results}
\label{sec:simulation-results}

\paragraph{Projection diagnostics.}

\begin{figure}[t]
 \centering
 \includegraphics[width=\linewidth]
 {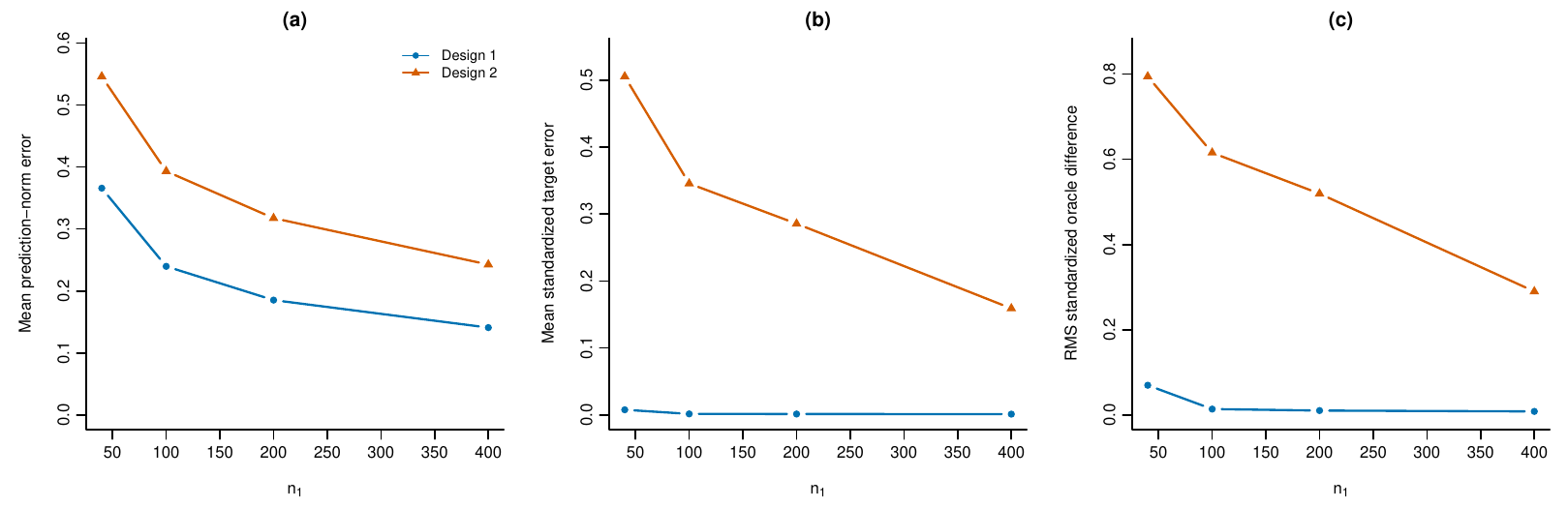}
 \caption{
 Direction learning and oracle agreement
 in Designs 1--2 with \(p=2000\) and
 \(\rho_0=0\).
 We use \(n\in\{200,500,1000,2000\}\),
 \(n_1=\lfloor0.2n\rfloor\), and
 \(R=500\) replications per configuration.
 Panels (a)--(c) report \(\bar d\),
 \(\bar t\), and \(g_{\mathrm{RMS}}\),
 respectively.
 Feasible and oracle estimates use the
 same evaluation observations within
 each replication.
 }
 \label{fig:direction-learning}
\end{figure}

Figure~\ref{fig:direction-learning}
shows that the prediction-norm direction
error decreases in both designs as
\(n_1\) increases from \(40\) to \(400\),
with larger errors in Design 2.
In Design 1, both \(\bar t\) and
\(g_{\mathrm{RMS}}\) remain small relative
to the oracle sampling scale.
In Design 2, \(\bar t\) decreases from
approximately \(0.51\) to \(0.16\), while
\(g_{\mathrm{RMS}}\) decreases from
approximately \(0.80\) to \(0.29\).
These trends indicate improved direction
learning and closer oracle agreement
as sample size increases, although
a finite-sample learning discrepancy
remains in Design 2.

\begin{figure}[t]
 \centering
 \includegraphics[width=0.85\linewidth]
 {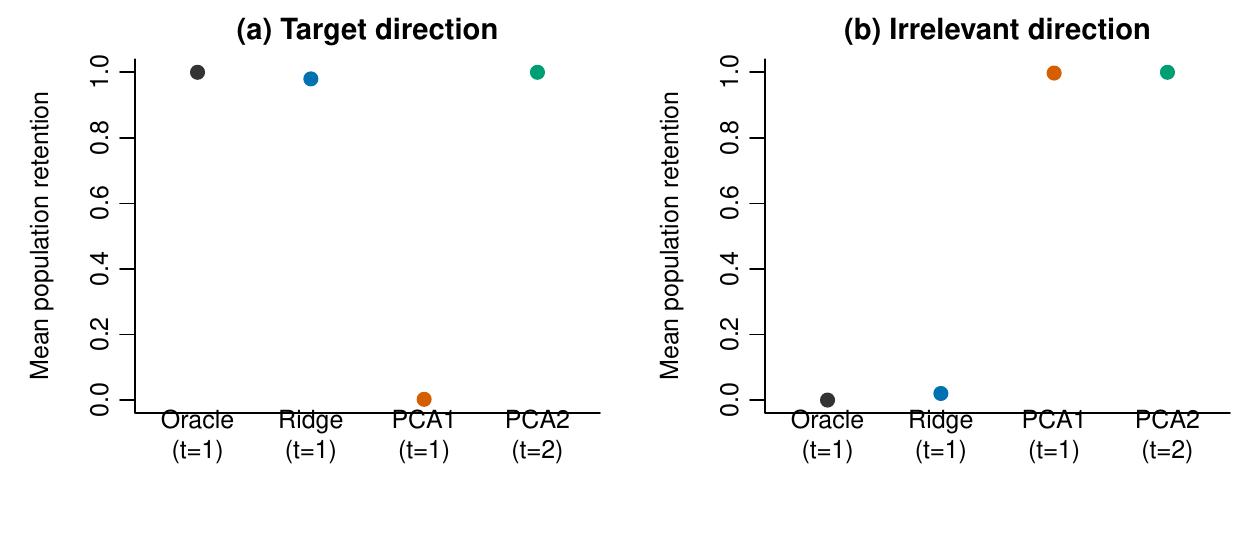}
 \caption{
 Target and irrelevant-direction retention
 in Design 2 with \(n=p=2000\),
 \(n_1=400\), and \(\rho_0=0\).
 Points show Monte Carlo averages over
 \(R=500\) replications.
 Oracle, Ridge, and PCA1 use one-dimensional
 projections, whereas PCA2 uses a
 two-dimensional projection.
 }
 \label{fig:retention}
\end{figure}

Figure~\ref{fig:retention} compares
population retention in Design 2.
Using a single direction, Ridge retains
approximately \(98\%\) of the target-index
variance and only \(2\%\) of the
irrelevant-score variance, approaching
the Oracle retention pattern.
PCA1 captures almost all of the stronger
irrelevant component and little of the
target component.
PCA2 retains nearly all of both components
through its two-dimensional projection.
These results illustrate the ability
of the learned one-dimensional reduction
to retain the target-relevant component
in the presence of stronger irrelevant
variation.

\paragraph{Testing performance.}
\begin{table}[t]
\centering
\caption{
Testing performance in Design 1.
Results are based on \(R=500\) Monte Carlo
replications per configuration at nominal
level \(\alpha=0.05\).
}
\label{tab:testing-design1}

\begingroup
\setlength{\tabcolsep}{4pt}
\renewcommand{\arraystretch}{1.15}

\begin{tabular*}{\linewidth}
{@{\extracolsep{\fill}}rr*{8}{r}@{}}
\toprule
& &
\multicolumn{4}{c}{Rejection rate}
&
\multicolumn{4}{c}{Total time (s)}
\\
\cmidrule(lr){3-6}
\cmidrule(lr){7-10}
\(n\) & \(p\)
& OPT & Oracle & P-DCov & RSZZ
& OPT & Oracle & P-DCov & RSZZ
\\
\midrule

\multicolumn{10}{l}{
Panel A: \(\rho_0=0\)
}
\\
\addlinespace[3pt]
 500 &  500
 & 0.044 & 0.046 & 0.050 & 0.064
 & 8.881 & 2.129 & 3369.317 & 217.253 \\
 500 & 1000
 & 0.048 & 0.050 & 0.042 & 0.044
 & 14.524 & 3.765 & 3058.803 & 495.615 \\
1000 &  500
 & 0.042 & 0.042 & 0.038 & 0.042
 & 25.292 & 4.572 & 8241.030 & 747.896 \\
1000 & 1000
 & 0.026 & 0.028 & 0.048 & 0.048
 & 46.500 & 10.225 & 20687.318 & 1914.479 \\

\addlinespace[7pt]
\multicolumn{10}{l}{
Panel B: \(\rho_0=-0.2\)
}
\\
\addlinespace[3pt]
 500 &  500
 & 0.988 & 0.988 & 0.980 & 0.994
 & 8.893 & 2.402 & 3354.083 & 221.024 \\
 500 & 1000
 & 0.986 & 0.986 & 0.992 & 0.996
 & 14.360 & 3.610 & 3059.717 & 494.999 \\
1000 &  500
 & 0.998 & 0.998 & 0.998 & 1.000
 & 23.707 & 5.108 & 8261.868 & 725.662 \\
1000 & 1000
 & 1.000 & 1.000 & 1.000 & 1.000
 & 46.225 & 9.102 & 20654.884 & 1913.294 \\

\addlinespace[7pt]
\multicolumn{10}{l}{
Panel C: \(\rho_0=0.2\)
}
\\
\addlinespace[3pt]
 500 &  500
 & 0.992 & 0.992 & 0.986 & 0.992
 & 9.313 & 2.428 & 3375.641 & 215.847 \\
 500 & 1000
 & 0.984 & 0.984 & 0.984 & 0.988
 & 14.558 & 3.754 & 3055.761 & 498.324 \\
1000 &  500
 & 1.000 & 1.000 & 1.000 & 1.000
 & 25.302 & 5.428 & 8234.129 & 719.160 \\
1000 & 1000
 & 1.000 & 1.000 & 1.000 & 1.000
 & 45.103 & 9.216 & 20701.388 & 1917.437 \\

\bottomrule
\end{tabular*}

\par\medskip
\begin{minipage}{\linewidth}
\footnotesize
Rejection rates estimate type-I error
when \(\rho_0=0\) and power otherwise.
Power is interpreted together with null
rejection rates.
OPT learns its direction from the training
sample and tests on the independent
evaluation sample.
Oracle uses the population target-preserving
direction and exactly the same evaluation
observations as OPT.
P-DCov and RSZZ use all \(n\) observations.
Total time sums the elapsed computation
times over all replications for each
method and configuration.
For Oracle, this excludes direction learning.
\end{minipage}

\endgroup
\end{table}

\paragraph{Testing performance in Design 1.}

Table~\ref{tab:testing-design1} shows that
Oracle's empirical null rejection rates
remain near \(0.05\), ranging from
\(0.044\) to \(0.058\).
Power approaches one as \(n\) increases
from \(500\) to \(1000\), while increasing
\(p\) slightly reduces power at \(n=500\). OPT closely follows Oracle, with rejection
rates differing by at most \(0.004\).
Its null rejection rates range from
\(0.046\) to \(0.054\), and its power
exhibits the same sample-size and
dimension patterns. P-DCov achieves high power, although
its empirical null rejection rates
slightly exceed \(0.05\) in every
configuration, ranging from \(0.052\)
to \(0.058\).
RSZZ also achieves high power, with
null rejection rates ranging from
\(0.042\) to \(0.070\).
Both attain power one when \(n=1000\). Oracle has the lowest computation times,
which exclude direction learning.
Under the reported implementations,
OPT is the fastest feasible method,
followed by RSZZ and P-DCov.

\paragraph{Testing performance in Designs 1--2.}

Under the Gaussian model and rank-one
alignment, the oracle version of our
projected test preserves the original
null and admits exact finite-sample
\(F\) calibration.
Tables~\ref{tab:testing-design1}
and~\ref{tab:testing-design2} show that
Oracle also achieves high power,
ranging from \(0.958\) to \(0.978\)
when \(n=500\).
Its population projection direction
is unknown in applications, motivating
OPT as a feasible implementation
that learns this direction from
training data.

OPT closely follows Oracle in Design 1,
with rejection rates differing by
at most \(0.002\).
Design 2 presents greater difficulty
for direction learning:
OPT exhibits size inflation when
\(n=500\), followed by improved
calibration at \(n=1000\).
At fixed \(p=1000\), its null rejection
rate decreases from \(0.114\) to \(0.050\),
and the gap from Oracle narrows
from \(0.068\) to \(0.008\).
These results show improved agreement
with the oracle benchmark as more
observations become available.

P-DCov achieves high power, although
its empirical type-I error rate
occasionally exceeds the nominal
significance level. RSZZ often has high alternative rejection
rates, accompanied by higher null
rejection rates, particularly in Design 2.
Under the reported implementations,
OPT requires substantially less
computation time than either competitor.

\paragraph{Testing performance in Design 3.}

The target-preserving Oracle combines
null rejection rates of \(0.042\)--\(0.052\)
with high power and the lowest reported
computation times.
Its power ranges from \(0.952\) to
\(0.970\) when \(n=500\).
These results illustrate the effectiveness
of low-dimensional testing when a
target-preserving population projection
is available.
The Oracle timing excludes direction
learning.

OPT has lower power, but its power
gap from Oracle narrows with increasing
sample size.
At fixed \(p=500\), the gap decreases
from \(0.374\) to \(0.064\) under
the negative alternative and from
\(0.456\) to \(0.180\) under the
positive alternative.
This shows improved agreement in power
with the oracle benchmark.
At \(p=1000\), however, its empirical
null rejection rates remain elevated,
at \(0.070\) and \(0.076\).

P-DCov and RSZZ achieve higher power
than OPT in the reported alternative
configurations.
Their empirical null rejection rates
occasionally exceed the nominal level,
ranging from \(0.056\) to \(0.060\)
for P-DCov and from \(0.044\) to
\(0.062\) for RSZZ.
Both require substantially more
computation time than OPT, which
remains the fastest feasible test
under the reported implementations.

\begin{table}[t]
\centering
\caption{
Testing performance in Design 2.
Results are based on \(R=500\) Monte Carlo
replications per configuration at nominal
level \(\alpha=0.05\).
}
\label{tab:testing-design2}

\begingroup
\setlength{\tabcolsep}{4pt}
\renewcommand{\arraystretch}{1.15}

\begin{tabular*}{\linewidth}
{@{\extracolsep{\fill}}rr*{8}{r}@{}}
\toprule
& &
\multicolumn{4}{c}{Rejection rate}
&
\multicolumn{4}{c}{Total time (s)}
\\
\cmidrule(lr){3-6}
\cmidrule(lr){7-10}
\(n\) & \(p\)
& OPT & Oracle & P-DCov & RSZZ
& OPT & Oracle & P-DCov & RSZZ
\\
\midrule

\multicolumn{10}{l}{
Panel A: \(\rho_0=0\)
}
\\
\addlinespace[3pt]
 500 &  500
 & 0.126 & 0.042 & 0.040 & 0.160
 & 11.618 & 1.634 & 3149.778 & 646.480 \\
 500 & 1000
 & 0.114 & 0.046 & 0.046 & 0.140
 & 18.897 & 4.081 & 2747.123 & 14512.820 \\
1000 &  500
 & 0.072 & 0.054 & 0.048 & 0.090
 & 35.986 & 3.427 & 9548.642 & 1965.254 \\
1000 & 1000
 & 0.050 & 0.042 & 0.046 & 0.090
 & 76.152 & 8.946 & 14291.398 & 189048.623 \\

\addlinespace[7pt]
\multicolumn{10}{l}{
Panel B: \(\rho_0=-0.2\)
}
\\
\addlinespace[3pt]
 500 &  500
 & 0.982 & 0.974 & 0.978 & 1.000
 & 11.263 & 1.695 & 3142.025 & 648.300 \\
 500 & 1000
 & 0.978 & 0.960 & 0.984 & 1.000
 & 20.395 & 3.822 & 3104.197 & 32805.780 \\
1000 &  500
 & 1.000 & 1.000 & 1.000 & 1.000
 & 36.518 & 3.580 & 9564.714 & 2019.580\\
1000 & 1000
 & 1.000 & 1.000 & 1.000 & 0.998
 & 81.899 & 9.471 & 19222.025 & 222181.677 \\

\addlinespace[7pt]
\multicolumn{10}{l}{
Panel C: \(\rho_0=0.2\)
}
\\
\addlinespace[3pt]
 500 &  500
 & 0.882 & 0.964 & 0.968 & 0.918
 & 11.882 & 1.798 & 3134.184 & 650.556 \\
 500 & 1000
 & 0.912 & 0.972 & 0.980 & 0.928
 & 21.266 & 4.693 & 3399.322 & 17152.098 \\
1000 &  500
 & 1.000 & 1.000 & 1.000 & 1.000
 & 35.666 & 3.505 & 9554.351 & 2008.910 \\
1000 & 1000
 & 0.998 & 1.000 & 1.000 & 0.996
 & 70.281 & 9.365 & 19016.484 & 226759.874 \\

\bottomrule
\end{tabular*}

\par\medskip
\begin{minipage}{\linewidth}

\end{minipage}

\endgroup
\end{table}

\begin{table}[t]
\centering
\caption{
Testing performance in Design 3.
Results are based on \(R=500\) Monte Carlo
replications per configuration at nominal
level \(\alpha=0.05\).
}
\label{tab:testing-design3}

\begingroup
\setlength{\tabcolsep}{4pt}
\renewcommand{\arraystretch}{1.15}

\begin{tabular*}{\linewidth}
{@{\extracolsep{\fill}}rr*{8}{r}@{}}
\toprule
& &
\multicolumn{4}{c}{Rejection rate}
&
\multicolumn{4}{c}{Total time (s)}
\\
\cmidrule(lr){3-6}
\cmidrule(lr){7-10}
\(n\) & \(p\)
& OPT & Oracle & P-DCov & RSZZ
& OPT & Oracle & P-DCov & RSZZ
\\
\midrule

\multicolumn{10}{l}{
Panel A: \(\rho_0=0\)
}
\\
\addlinespace[3pt]
 500 &  500
 & 0.048 & 0.042 & 0.058 & 0.044
 & 11.925 & 2.556 & 2190.258 & 18129.597 \\
 500 & 1000
 & 0.070 & 0.052 & 0.056 & 0.062
 & 19.353 & 3.999 & 2183.280 & 14729.668 \\
1000 &  500
 & 0.050 & 0.052 & 0.060 & 0.048
 & 32.500 & 3.910 & 3664.589 & 9346.464 \\
1000 & 1000
 & 0.076 & 0.044 & 0.060 & 0.058
 & 78.270 & 8.805 & 12324.448 & 196219.650 \\

\addlinespace[7pt]
\multicolumn{10}{l}{
Panel B: \(\rho_0=-0.2\)
}
\\
\addlinespace[3pt]
 500 &  500
 & 0.588 & 0.962 & 0.882 & 0.858
 & 11.149 & 2.266 & 1870.971 & 18645.181 \\
 500 & 1000
 & 0.586 & 0.970 & 0.834 & 0.810
 & 18.956 & 4.198 & 2294.596 & 15036.593 \\
1000 &  500
 & 0.936 & 1.000 & 0.998 & 1.000
 & 31.794 & 3.757 & 3526.567 & 9077.059\\
1000 & 1000
 & 0.880 & 1.000 & 0.998 & 0.998
 & 78.499 & 9.631 & 11955.163 & 191603.465 \\

\addlinespace[7pt]
\multicolumn{10}{l}{
Panel C: \(\rho_0=0.2\)
}
\\
\addlinespace[3pt]
 500 &  500
 & 0.496 & 0.952 & 0.886 & 0.716
 & 12.097 & 3.011 & 2111.211 & 18029.751 \\
 500 & 1000
 & 0.388 & 0.958 & 0,762 & 0.662
 & 18.825 & 4.219 & 2127.352 & 14347.664 \\
1000 &  500
 & 0.820 & 1.000 & 1.000 & 0.998
 & 32.561 & 3.952 & 3880.299 & 9895.238 \\
1000 & 1000
 & 0.764 & 0.998 & 1.000 & 0.992
 & 89.298 & 11.215 & 13771.623 & 245607.899 \\

\bottomrule
\end{tabular*}

\par\medskip
\begin{minipage}{\linewidth}
\end{minipage}

\endgroup
\end{table}

\begin{table}[t]
\centering
\caption{
Testing performance in Design 4.
Results are based on \(R=500\) Monte Carlo
replications per configuration at nominal
level \(\alpha=0.05\).
}
\label{tab:testing-design3}

\begingroup
\setlength{\tabcolsep}{4pt}
\renewcommand{\arraystretch}{1.15}

\begin{tabular*}{\linewidth}
{@{\extracolsep{\fill}}rr*{8}{r}@{}}
\toprule
& &
\multicolumn{4}{c}{Rejection rate}
&
\multicolumn{4}{c}{Total time (s)}
\\
\cmidrule(lr){3-6}
\cmidrule(lr){7-10}
\(n\) & \(p\)
& OPT & Oracle & P-DCov & RSZZ
& OPT & Oracle & P-DCov & RSZZ
\\
\midrule

\multicolumn{10}{l}{
Panel A: \(\rho_0=0\)
}
\\
\addlinespace[3pt]
 500 &  8000
 & 0.052 & 0.036 & 0.348 & 0.932
 & 324.382 & 32.376 & 14737.990 & 5595.125 \\
 500 & 10000
 & 0.044 & 0.050 & 0.292 & 0.920
 & 345.851 & 34.792 & 15068.499 & 6484.351 \\
1000 &  500
 & 0.050 & 0.052 & 0.060 & 0.048
 & 32.500 & 3.910 & 3664.589 & 9346.464 \\
1000 & 1000
 & 0.076 & 0.044 & 0.060 & 0.058
 & 78.270 & 8.805 & 12324.448 & 196219.650 \\

\addlinespace[7pt]
\multicolumn{10}{l}{
Panel B: \(\rho_0=-0.2\)
}
\\
\addlinespace[3pt]
 500 &  8000
 & 0.984 & 0.960 & 0.802 & 1.000
 & 324.830 & 32.315 & 14729.039 & 5507.803 \\
 500 & 10000
 & 0.966 & 0.948 & 0.748 & 1.000
 & 345.733 & 35.655 & 15047.873 & 6370.851 \\
1000 &  500
 & 0.936 & 1.000 & 0.998 & 1.000
 & 31.794 & 3.757 & 3526.567 & 9077.059\\
1000 & 1000
 & 0.880 & 1.000 & 0.998 & 0.998
 & 78.499 & 9.631 & 11955.163 & 191603.465 \\

\addlinespace[7pt]
\multicolumn{10}{l}{
Panel C: \(\rho_0=0.2\)
}
\\
\addlinespace[3pt]
 500 &  8000
 & 0.920 & 0.958 & 0.560 & 0.074
 & 321.347 & 31.981 & 14749.783 & 5619.086 \\
 500 & 10000
 & 0.952 & 0.976 & 0.554 & 0.068
 & 343.160 & 34.755 & 15001.498 & 6425.402 \\
1000 &  500
 & 0.820 & 1.000 & 1.000 & 0.998
 & 32.561 & 3.952 & 3880.299 & 9895.238 \\
1000 & 1000
 & 0.764 & 0.998 & 1.000 & 0.992
 & 89.298 & 11.215 & 13771.623 & 245607.899 \\

\bottomrule
\end{tabular*}

\par\medskip
\begin{minipage}{\linewidth}
\end{minipage}

\endgroup
\end{table}

\begin{table}[t]
\centering
\caption{
Rejection rates and cumulative computation
times in Design 4 with \(n=500\) and
\(p=10000\).
Results are based on \(R=500\) replications
per configuration at nominal level
\(\alpha=0.05\).
}
\label{tab:design4-testing}

\begingroup
\renewcommand{\arraystretch}{1.15}

\begin{tabular*}{0.85\linewidth}
{@{\extracolsep{\fill}}lrr@{}}
\toprule
Method & Rejection rate & Total time (s) \\
\midrule

\multicolumn{3}{l}{Panel A: \(\rho_0=0\)} \\
\addlinespace[3pt]
OPT    & 0.068 &    427.686 \\
Oracle & 0.052 &     38.881 \\
P-DCov & 0.284 & 18\,511.321 \\
RSZZ   & 0.946 &  7\,520.791 \\

\addlinespace[7pt]
\multicolumn{3}{l}{Panel B: \(\rho_0=-0.2\)} \\
\addlinespace[3pt]
OPT    & 0.976 &    430.739 \\
Oracle & 0.966 &     40.786 \\
P-DCov & 0.740 & 18\,539.182 \\
RSZZ   & 1.000 &  7\,438.209 \\

\addlinespace[7pt]
\multicolumn{3}{l}{Panel C: \(\rho_0=0.2\)} \\
\addlinespace[3pt]
OPT    & 0.968 &    431.517 \\
Oracle & 0.986 &     39.171 \\
P-DCov & 0.578 & 18\,511.708 \\
RSZZ   & 0.056 &  7\,443.327 \\

\bottomrule
\end{tabular*}

\par\medskip

\endgroup
\end{table}

\section{Discussion}
\label{s5} 
This paper develops a target-specific reduction
framework for partial correlation inference
in high-dimensional Gaussian models.
The central idea is to replace the
high-dimensional conditioning vector with
a low-dimensional projection that preserves
the inferential target under explicit
structural conditions. This approach
provides an alternative to coordinate-sparsity
assumptions without requiring recovery of
the full covariance or precision matrix.

The theoretical guarantees depend on two
distinct requirements: population target
preservation and statistical learnability.
Rank-one alignment ensures that the
population optimal direction preserves
the original partial correlation, while
regularized effective-dimension and
target-specific bias conditions control
the estimation error of its feasible
ridge-based counterpart. Their combination
yields oracle-equivalent estimation and
asymptotically valid testing for the
original null. An explicit coordinate-dense
model demonstrates that these conditions
can hold even when the ambient dimension
exceeds the sample size.

These guarantees do not extend to
unrestricted dense high-dimensional models.
In particular, rank-one alignment restricts
the relevant population regression directions
to a common one-dimensional subspace, and
the learning conditions require sufficient
covariance structure to estimate that
subspace at the inference scale.
Extensions to higher-dimensional
target-preserving subspaces, data-adaptive
regularization, and inference beyond
Gaussian models are promising directions
for future research.

\section*{Appendices}
In the supplemental material, we provide proofs of the theoretical results, details on the choice of tuning parameters, and additional simulation results under alternative covariance structures referenced in Section \ref{s4}.

\acks{The first author gratefully acknowledges the guidance received from Professor Hansheng Wang during a research visit to Peking University. The authors also thank Professor Ziqi Chen (East China Normal University) for valuable suggestions that substantially improved the paper.

This work was supported by the Guangxi Natural Science Foundation (Grant No. 2025GXNSFAA069842) and the National Natural Science Foundation of China (Grant Nos. 11801105 and 72171234).

The authors declare no competing interests.}

\newpage
\clearpage

\bibliography{gpetcite}

@article{buhlmann2010,
  author = {B{\"u}hlmann, Peter and Kalisch, Markus and Maathuis, Marloes H.},
  title = {Variable selection in high-dimensional linear models: Partially faithful distributions and the {PC}-simple algorithm},
  journal = {Biometrika},
  volume = {97},
  number = {2},
  pages = {261--278},
  year = {2010},
  publisher = {Oxford University Press}
}

@article{cai2011constrained,
  author = {Cai, T. Tony and Liu, Weidong and Luo, Xi},
  title = {A constrained $\ell_1$ minimization approach to sparse precision matrix estimation},
  journal = {Journal of the American Statistical Association},
  year = {2011},
  volume = {106},
  number = {494},
  pages = {594--607}
}

@article{fan2020projection,
  author = {Fan, Jianqing and Feng, Yang and Xia, Lucy},
  title = {A projection-based conditional dependence measure with applications to high-dimensional undirected graphical models},
  journal = {Journal of Econometrics},
  year = {2020},
  volume = {218},
  number = {1},
  pages = {119--139}
}

@article{friedman_hastie_tibshirani_2008,
  author = {Friedman, Jerome and Hastie, Trevor and Tibshirani, Robert},
  title = {Sparse inverse covariance estimation with the graphical {LASSO}},
  journal = {Biostatistics},
  year = {2008},
  volume = {9},
  number = {3},
  pages = {432--441}
}

@article{khare_oh_rajaratnam_2015,
  author = {Khare, Kshitij and Oh, Sang-Yun and Rajaratnam, Bala},
  title = {A convex pseudolikelihood framework for high dimensional partial correlation estimation with convergence guarantees},
  journal = {Journal of the Royal Statistical Society Series B: Statistical Methodology},
  volume = {77},
  number = {4},
  pages = {803--825},
  year = {2015}
}

@article{LW2017,
  author = {Ledoit, Olivier and Wolf, Michael},
  title = {Nonlinear shrinkage of the covariance matrix for portfolio selection: {Markowitz} meets {Goldilocks}},
  journal = {The Review of Financial Studies},
  volume = {30},
  number = {12},
  pages = {4349--4388},
  year = {2017}
}

@article{li2022linear,
  author = {Li, Changcheng and Li, Runze},
  title = {Linear hypothesis testing in linear models with high-dimensional responses},
  journal = {Journal of the American Statistical Association},
  year = {2022},
  volume = {117},
  number = {540},
  pages = {1738--1750}
}

@article{li2017,
  author = {Li, Runze and Liu, Jingyuan and Lou, Lejia},
  title = {Variable selection via partial correlation},
  journal = {Statistica Sinica},
  volume = {27},
  number = {3},
  pages = {983--1007}, 
  year = {2017}
}

@article{liang2015equivalent,
  author = {Liang, Faming and Song, Qifan and Qiu, Peihua},
  title = {An equivalent measure of partial correlation coefficients for high-dimensional {G}aussian graphical models},
  journal = {Journal of the American Statistical Association},
  volume = {110},
  number = {511},
  pages = {1248--1265},
  year = {2015},
  publisher = {Taylor \& Francis}
}

@article{luo2024functional,
  title = {Functional connectivity development along the sensorimotor-association axis enhances the cortical hierarchy},
  author = {Luo, Audrey C. and Sydnor, Valerie J. and Pines, Adam and Larsen, Bart and Alexander-Bloch, Aaron F. and Cieslak, Matthew and Covitz, Sydney and Chen, Andrew A. and Esper, Nathalia Bianchini and Feczko, Eric and others},
  journal = {Nature Communications},
  volume = {15},
  number = {1},
  pages = {3511},
  year = {2024}
}

@article{meinshausen_buhlmann_2006,
  author = {Meinshausen, Nicolai and B{\"u}hlmann, Peter},
  title = {High-dimensional graphs and variable selection with the {LASSO}},
  journal = {The Annals of Statistics},
  year = {2006},
  volume = {34},
  number = {3},
  pages = {1436--1462}
}

@article{peng2009,
  author = {Peng, Jie and Wang, Pei and Zhou, Nengfeng and Zhu, Ji},
  title = {Partial correlation estimation by joint sparse regression models},
  journal = {Journal of the American Statistical Association},
  volume = {104},
  number = {486},
  pages = {735--746},
  year = {2009},
  publisher = {Taylor \& Francis}
}

@article{scalreg2013,
  author = {Sun, Tingni and Zhang, Cun-Hui},
  title = {Sparse matrix inversion with scaled {LASSO}},
  journal = {Journal of Machine Learning Research},
  year = {2013},
  volume = {14},
  number = {106},
  pages = {3385--3418}
}

@article{huang2022kernel,
  title={Kernel partial correlation coefficient---a measure of conditional dependence},
  author={Huang, Zhen and Deb, Nabarun and Sen, Bodhisattva},
  journal={Journal of Machine Learning Research},
  volume={23},
  number={216},
  pages={1--58},
  year={2022}
}

@article{azadkia2021simple,
  title={A simple measure of conditional dependence},
  author={Azadkia, Mona and Chatterjee, Sourav},
  journal={The Annals of Statistics},
  volume={49},
  number={6},
  pages={3070--3102},
  year={2021},
  publisher={JSTOR}
}

@article{ren2015asymptotic,
  author  = {Ren, Zhao and Sun, Tingni and Zhang, Cun-Hui
             and Zhou, Harrison H.},
  title   = {Asymptotic Normality and Optimalities in Estimation
             of Large Gaussian Graphical Models},
  journal = {The Annals of Statistics},
  year    = {2015},
  volume  = {43},
  number  = {3},
  pages   = {991--1026},
  doi     = {10.1214/14-AOS1286}
}

@book{muirhead1982aspects,
  author    = {Muirhead, Robb J.},
  title     = {Aspects of Multivariate Statistical Theory},
  publisher = {John Wiley \& Sons},
  year      = {1982}
}

@article{hsu2014random,
  author  = {Hsu, Daniel and Kakade, Sham M. and Zhang, Tong},
  title   = {Random Design Analysis of Ridge Regression},
  journal = {Foundations of Computational Mathematics},
  year    = {2014},
  volume  = {14},
  pages   = {569--600},
  doi     = {10.1007/s10208-014-9192-1}
}

@article{caponnetto2007optimal,
  author  = {Caponnetto, Andrea and De Vito, Ernesto},
  title   = {Optimal Rates for the Regularized
             Least-Squares Algorithm},
  journal = {Foundations of Computational Mathematics},
  year    = {2007},
  volume  = {7},
  number  = {3},
  pages   = {331--368},
  doi     = {10.1007/s10208-006-0196-8}
}

@article{chernozhukov2018double,
  author  = {Chernozhukov, Victor and
             Chetverikov, Denis and
             Demirer, Mert and
             Duflo, Esther and
             Hansen, Christian and
             Newey, Whitney and
             Robins, James},
  title   = {Double/Debiased Machine Learning for
             Treatment and Structural Parameters},
  journal = {The Econometrics Journal},
  year    = {2018},
  volume  = {21},
  number  = {1},
  pages   = {C1--C68},
  doi     = {10.1111/ectj.12097}
}

\end{document}